
%
\expandafter\ifx\csname phyzzx\endcsname\relax
 \message{It is better to use PHYZZX format than to
          \string\input\space PHYZZX}\else
 \wlog{PHYZZX macros are already loaded and are not
          \string\input\space again}%
 \endinput \fi
\catcode`\@=11 
\let\rel@x=\relax
\let\n@expand=\relax
\def\pr@tect{\let\n@expand=\noexpand}
\let\protect=\pr@tect
\let\gl@bal=\global
%
%
%
\newfam\cpfam
\newdimen\b@gheight             \b@gheight=12pt
\newcount\f@ntkey               \f@ntkey=0
\def\f@m{\afterassignment\samef@nt\f@ntkey=}
\def\samef@nt{\fam=\f@ntkey\the\textfont\f@ntkey\rel@x}
\def\setstr@t{\setbox\strutbox=\hbox{\vrule height 0.85\b@gheight
                                depth 0.35\b@gheight width\z@ }}
\input phyzzx.fonts
%
\def\rm{\n@expand\f@m0 }
\def\mit{\n@expand\f@m1 }         \let\oldstyle=\mit
\def\cal{\n@expand\f@m2}
\def\it{\n@expand\f@m\itfam}
\def\sl{\n@expand\f@m\slfam}
\def\bf{\n@expand\f@m\bffam}
\def\tt{\n@expand\f@m\ttfam}
\def\caps{\n@expand\f@m\cpfam}    \let\cp=\caps
\def\em@{\rel@x\ifnum\f@ntkey=0\it\else
        \ifnum\f@ntkey=\bffam\it\else\rm\fi \fi }
\def\em{\n@expand\em@}
\def\fourteenpoint{\fourteenf@nts \samef@nt \b@gheight=14pt \setstr@t }
\def\twelvepoint{\twelvef@nts \samef@nt \b@gheight=12pt \setstr@t }
\def\tenpoint{\tenf@nts \samef@nt \b@gheight=10pt \setstr@t }
\normalbaselineskip = 20pt plus 0.2pt minus 0.1pt
\normallineskip = 1.5pt plus 0.1pt minus 0.1pt
\normallineskiplimit = 1.5pt
\newskip\normaldisplayskip
\normaldisplayskip = 20pt plus 5pt minus 10pt
\newskip\normaldispshortskip
\normaldispshortskip = 6pt plus 5pt
\newskip\normalparskip
\normalparskip = 6pt plus 2pt minus 1pt
\newskip\skipregister
\skipregister = 5pt plus 2pt minus 1.5pt
\newif\ifsingl@
\newif\ifdoubl@
\newif\iftwelv@  \twelv@true
\def\singlespace{\singl@true\doubl@false\spaces@t}
\def\doublespace{\singl@false\doubl@true\spaces@t}
\def\normalspace{\singl@false\doubl@false\spaces@t}
\def\Tenpoint{\tenpoint\twelv@false\spaces@t}
\def\Twelvepoint{\twelvepoint\twelv@true\spaces@t}
\def\spaces@t{\rel@x
      \iftwelv@ \ifsingl@\subspaces@t3:4;\else\subspaces@t1:1;\fi
       \else \ifsingl@\subspaces@t3:5;\else\subspaces@t4:5;\fi \fi
      \ifdoubl@ \multiply\baselineskip by 5
         \divide\baselineskip by 4 \fi }
\def\subspaces@t#1:#2;{
      \baselineskip = \normalbaselineskip
      \multiply\baselineskip by #1 \divide\baselineskip by #2
      \lineskip = \normallineskip
      \multiply\lineskip by #1 \divide\lineskip by #2
      \lineskiplimit = \normallineskiplimit
      \multiply\lineskiplimit by #1 \divide\lineskiplimit by #2
      \parskip = \normalparskip
      \multiply\parskip by #1 \divide\parskip by #2
      \abovedisplayskip = \normaldisplayskip
      \multiply\abovedisplayskip by #1 \divide\abovedisplayskip by #2
      \belowdisplayskip = \abovedisplayskip
      \abovedisplayshortskip = \normaldispshortskip
      \multiply\abovedisplayshortskip by #1
        \divide\abovedisplayshortskip by #2
      \belowdisplayshortskip = \abovedisplayshortskip
      \advance\belowdisplayshortskip by \belowdisplayskip
      \divide\belowdisplayshortskip by 2
      \smallskipamount = \skipregister
      \multiply\smallskipamount by #1 \divide\smallskipamount by #2
      \medskipamount = \smallskipamount \multiply\medskipamount by 2
      \bigskipamount = \smallskipamount \multiply\bigskipamount by 4 }
\def\normalbaselines{ \baselineskip=\normalbaselineskip
   \lineskip=\normallineskip \lineskiplimit=\normallineskip
   \iftwelv@\else \multiply\baselineskip by 4 \divide\baselineskip by 5
     \multiply\lineskiplimit by 4 \divide\lineskiplimit by 5
     \multiply\lineskip by 4 \divide\lineskip by 5 \fi }
\Twelvepoint  
\interlinepenalty=50
\interfootnotelinepenalty=5000
\predisplaypenalty=9000
\postdisplaypenalty=500
\hfuzz=1pt
\vfuzz=0.2pt
\newdimen\HOFFSET  \HOFFSET=0pt
\newdimen\VOFFSET  \VOFFSET=0pt
\newdimen\HSWING   \HSWING=0pt
\dimen\footins=8in
%
%
%
\newskip\pagebottomfiller
\pagebottomfiller=\z@ plus \z@ minus \z@
\def\pagecontents{
   \ifvoid\topins\else\unvbox\topins\vskip\skip\topins\fi
   \dimen@ = \dp255 \unvbox255
   \vskip\pagebottomfiller
   \ifvoid\footins\else\vskip\skip\footins\footrule\unvbox\footins\fi
   \ifr@ggedbottom \kern-\dimen@ \vfil \fi }
\def\makeheadline{\vbox to 0pt{ \skip@=\topskip
      \advance\skip@ by -12pt \advance\skip@ by -2\normalbaselineskip
      \vskip\skip@ \line{\vbox to 12pt{}\the\headline} \vss
      }\nointerlineskip}
\def\makefootline{\baselineskip = 1.5\normalbaselineskip
                 \line{\the\footline}}
\newif\iffrontpage
\newif\ifp@genum
\def\nopagenumbers{\p@genumfalse}
\def\pagenumbers{\p@genumtrue}
\pagenumbers
\newtoks\paperheadline
\newtoks\paperfootline
\newtoks\letterheadline
\newtoks\letterfootline
\newtoks\letterinfo
\newtoks\date
\paperheadline={\hfil}
\paperfootline={\hss\iffrontpage\else\ifp@genum\tenrm\folio\hss\fi\fi}
\letterheadline{\iffrontpage \hfil \else
    \rm \ifp@genum page~~\folio\fi \hfil\the\date \fi}
\letterfootline={\iffrontpage\the\letterinfo\else\hfil\fi}
\letterinfo={\hfil}
\def\monthname{\rel@x\ifcase\month 0/\or January\or February\or
   March\or April\or May\or June\or July\or August\or September\or
   October\or November\or December\else\number\month/\fi}
\def\today{\monthname~\number\day, \number\year}
\date={\today}
\headline=\paperheadline 
\footline=\paperfootline 
\countdef\pageno=1      \countdef\pagen@=0
\countdef\pagenumber=1  \pagenumber=1
\def\advancepageno{\gl@bal\advance\pagen@ by 1
   \ifnum\pagenumber<0 \gl@bal\advance\pagenumber by -1
    \else\gl@bal\advance\pagenumber by 1 \fi
    \gl@bal\frontpagefalse  \swing@ }
\def\folio{\ifnum\pagenumber<0 \romannumeral-\pagenumber
           \else \number\pagenumber \fi }
\def\swing@{\ifodd\pagenumber \gl@bal\advance\hoffset by -\HSWING
             \else \gl@bal\advance\hoffset by \HSWING \fi }
\def\footrule{\dimen@=\prevdepth\nointerlineskip
   \vbox to 0pt{\vskip -0.25\baselineskip \hrule width 0.35\hsize \vss}
   \prevdepth=\dimen@ }
\let\footnotespecial=\rel@x
\newdimen\footindent
\footindent=24pt
\def\Textindent#1{\noindent\llap{#1\enspace}\ignorespaces}
\def\Vfootnote#1{\insert\footins\bgroup
   \interlinepenalty=\interfootnotelinepenalty \floatingpenalty=20000
   \singl@true\doubl@false\Tenpoint
   \splittopskip=\ht\strutbox \boxmaxdepth=\dp\strutbox
   \leftskip=\footindent \rightskip=\z@skip
   \parindent=0.5\footindent \parfillskip=0pt plus 1fil
   \spaceskip=\z@skip \xspaceskip=\z@skip \footnotespecial
   \Textindent{#1}\footstrut\futurelet\next\fo@t}

\def\vfootnote#1{\Vfootnote{${#1}$}}
\def\footnote#1{\attach{#1}\vfootnote{#1}}

\def\foot{\attach\footsymbolgen\vfootnote{\footsymbol}}
\let\footsymbol=\star
\newcount\lastf@@t           \lastf@@t=-1
\newcount\footsymbolcount    \footsymbolcount=0
\newif\ifPhysRev
\def\footsymbolgen{\bumpfootsymbolcount \generatefootsymbol \footsymbol }
\def\bumpfootsymbolcount{\rel@x
   \iffrontpage \bumpfootsymbolpos \else \advance\lastf@@t by 1
     \ifPhysRev \bumpfootsymbolneg \else \bumpfootsymbolpos \fi \fi
   \gl@bal\lastf@@t=\pagen@ }
\def\bumpfootsymbolpos{\ifnum\footsymbolcount <0
                            \gl@bal\footsymbolcount =0 \fi
    \ifnum\lastf@@t<\pagen@ \gl@bal\footsymbolcount=0
     \else \gl@bal\advance\footsymbolcount by 1 \fi }
\def\bumpfootsymbolneg{\ifnum\footsymbolcount >0
             \gl@bal\footsymbolcount =0 \fi
         \gl@bal\advance\footsymbolcount by -1 }
\def\fd@f#1 {\xdef\footsymbol{\mathchar"#1 }}
\def\generatefootsymbol{\ifcase\footsymbolcount \fd@f 13F \or \fd@f 279
        \or \fd@f 27A \or \fd@f 278 \or \fd@f 27B \else
        \ifnum\footsymbolcount <0 \fd@f{023 \number-\footsymbolcount }
         \else \fd@f 203 {\loop \ifnum\footsymbolcount >5
                \fd@f{203 \footsymbol } \advance\footsymbolcount by -1
                \repeat }\fi \fi }

\def\nonfrenchspacing{\sfcode`\.=3001 \sfcode`\!=3000 \sfcode`\?=3000
        \sfcode`\:=2000 \sfcode`\;=1500 \sfcode`\,=1251 }
\nonfrenchspacing
\newdimen\d@twidth
{\setbox0=\hbox{s.} \gl@bal\d@twidth=\wd0 \setbox0=\hbox{s}
        \gl@bal\advance\d@twidth by -\wd0 }
\def\removehglue{\loop \unskip \ifdim\lastskip >\z@ \repeat }
\def\roll@ver#1{\removehglue \nobreak \count255 =\spacefactor \dimen@=\z@
        \ifnum\count255 =3001 \dimen@=\d@twidth \fi
        \ifnum\count255 =1251 \dimen@=\d@twidth \fi
    \iftwelv@ \kern-\dimen@ \else \kern-0.83\dimen@ \fi
   #1\spacefactor=\count255 }
\def\step@ver#1{\rel@x \ifmmode #1\else \ifhmode
        \roll@ver{${}#1$}\else {\setbox0=\hbox{${}#1$}}\fi\fi }
\def\attach#1{\step@ver{\strut^{\mkern 2mu #1} }}
%
%
%
\newcount\chapternumber      \chapternumber=0
\newcount\sectionnumber      \sectionnumber=0
\newcount\equanumber         \equanumber=0
\let\chapterlabel=\rel@x
\let\sectionlabel=\rel@x
\newtoks\chapterstyle        \chapterstyle={\Number}
\newtoks\sectionstyle        \sectionstyle={\Number}
\newskip\chapterskip         \chapterskip=\bigskipamount
\newskip\sectionskip         \sectionskip=\medskipamount
\newskip\headskip            \headskip=8pt plus 3pt minus 3pt
\newdimen\chapterminspace    \chapterminspace=15pc
\newdimen\sectionminspace    \sectionminspace=10pc
\newdimen\referenceminspace  \referenceminspace=20pc
\newif\ifcn@                 \cn@true
\newif\ifcn@@                \cn@@false
\def\numberedchapters{\cn@true}
\def\unnumberedchapters{\cn@false\sequentialequations}
\def\chapterreset{\gl@bal\advance\chapternumber by 1
   \ifnum\equanumber<0 \else\gl@bal\equanumber=0\fi
   \sectionnumber=0 \let\sectionlabel=\rel@x
   \ifcn@ \gl@bal\cn@@true {\pr@tect
       \xdef\chapterlabel{\the\chapterstyle{\the\chapternumber}}}%
    \else \gl@bal\cn@@false \gdef\chapterlabel{\rel@x}\fi }
\def\@alpha#1{\count255='140 \advance\count255 by #1\char\count255}
 \def\alphabetic{\n@expand\@alpha}
\def\@Alpha#1{\count255='100 \advance\count255 by #1\char\count255}
 \def\Alphabetic{\n@expand\@Alpha}
\def\@Roman#1{\uppercase\expandafter{\romannumeral #1}}
 \def\Roman{\n@expand\@Roman}
\def\@roman#1{\romannumeral #1}    \def\roman{\n@expand\@roman}
\def\@number#1{\number #1}         \def\Number{\n@expand\@number}
\def\BLANK#1{\rel@x}               
\def\titleparagraphs{\interlinepenalty=9999
     \leftskip=0.03\hsize plus 0.22\hsize minus 0.03\hsize
     \rightskip=\leftskip \parfillskip=0pt
     \hyphenpenalty=9000 \exhyphenpenalty=9000
     \tolerance=9999 \pretolerance=9000
     \spaceskip=0.333em \xspaceskip=0.5em }
\def\titlestyle#1{\par\begingroup \titleparagraphs
     \iftwelv@\fourteenpoint\else\twelvepoint\fi
   \noindent #1\par\endgroup }
\def\spacecheck#1{\dimen@=\pagegoal\advance\dimen@ by -\pagetotal
   \ifdim\dimen@<#1 \ifdim\dimen@>0pt \vfil\break \fi\fi}
\def\chapter#1{\par \penalty-300 \vskip\chapterskip
   \spacecheck\chapterminspace
   \chapterreset \titlestyle{\ifcn@@\chapterlabel.~\fi #1}
   \nobreak\vskip\headskip \penalty 30000
   {\pr@tect\wlog{\string\chapter\space \chapterlabel}} }

\def\section#1{\par \ifnum\lastpenalty=30000\else
   \penalty-200\vskip\sectionskip \spacecheck\sectionminspace\fi
   \gl@bal\advance\sectionnumber by 1
   {\pr@tect
   \xdef\sectionlabel{\ifcn@@ \chapterlabel.\fi
       \the\sectionstyle{\the\sectionnumber}}%
   \wlog{\string\section\space \sectionlabel}}%
   \noindent {\caps\enspace\sectionlabel.~~#1}\par
   \nobreak\vskip\headskip \penalty 30000 }
\def\subsection#1{\par
   \ifnum\the\lastpenalty=30000\else \penalty-100\smallskip \fi
   \noindent\undertext{#1}\enspace \vadjust{\penalty5000}}

\def\undertext#1{\vtop{\hbox{#1}\kern 1pt \hrule}}
\def\APPENDIX#1#2{\par\penalty-300\vskip\chapterskip
   \spacecheck\chapterminspace \chapterreset \xdef\chapterlabel{#1}
   \titlestyle{APPENDIX #2} \nobreak\vskip\headskip \penalty 30000
   \wlog{\string\Appendix~\chapterlabel} }
\def\Appendix#1{\APPENDIX{#1}{#1}}
\def\appendix{\APPENDIX{A}{}}
%
%
%
%

\def\eqn{\eqno\eqname}

\def\eqinsert#1{\noalign{\dimen@=\prevdepth \nointerlineskip
   \setbox0=\hbox to\displaywidth{\hfil #1}
   \vbox to 0pt{\kern 0.5\baselineskip\hbox{$\!\box0\!$}\vss}
   \prevdepth=\dimen@}}
%

%
%
\def\GENITEM#1;#2{\par \hangafter=0 \hangindent=#1
    \Textindent{$ #2 $}\ignorespaces}
\outer\def\newitem#1=#2;{\gdef#1{\GENITEM #2;}}

\newdimen\itemsize                \itemsize=30pt
\newitem\item=1\itemsize;
\newitem\sitem=1.75\itemsize;     
\newitem\ssitem=2.5\itemsize;     
\outer\def\newlist#1=#2&#3&#4;{\toks0={#2}\toks1={#3}%
   \count255=\escapechar \escapechar=-1
   \alloc@0\list\countdef\insc@unt\listcount     \listcount=0
   \edef#1{\par
      \countdef\listcount=\the\allocationnumber
      \advance\listcount by 1
      \hangafter=0 \hangindent=#4
      \Textindent{\the\toks0{\listcount}\the\toks1}}
   \expandafter\expandafter\expandafter
    \edef\c@t#1{begin}{\par
      \countdef\listcount=\the\allocationnumber \listcount=1
      \hangafter=0 \hangindent=#4
      \Textindent{\the\toks0{\listcount}\the\toks1}}
   \expandafter\expandafter\expandafter
    \edef\c@t#1{con}{\par \hangafter=0 \hangindent=#4 \noindent}
   \escapechar=\count255}
\def\c@t#1#2{\csname\string#1#2\endcsname}
\newlist\point=\Number&.&1.0\itemsize;
\newlist\subpoint=(\alphabetic&)&1.75\itemsize;
\newlist\subsubpoint=(\roman&)&2.5\itemsize;
%

%
%
%
%
\newcount\referencecount     \referencecount=0
\newcount\lastrefsbegincount \lastrefsbegincount=0
\newif\ifreferenceopen       \newwrite\referencewrite
\newdimen\refindent          \refindent=30pt
\def\normalrefmark#1{\attach{\scriptscriptstyle [ #1 ] }}
\let\PRrefmark=\attach
\def\NPrefmark#1{\step@ver{{\;[#1]}}}
\def\refmark#1{\rel@x\ifPhysRev\PRrefmark{#1}\else\normalrefmark{#1}\fi}
\def\refend@{\refmark{\number\referencecount}}
\def\refend{\refend@{}\space }
\def\refsend{\refmark{\count255=\referencecount
   \advance\count255 by-\lastrefsbegincount
   \ifcase\count255 \number\referencecount
   \or \number\lastrefsbegincount,\number\referencecount
   \else \number\lastrefsbegincount-\number\referencecount \fi}\space }
\def\REFNUM#1{\rel@x \gl@bal\advance\referencecount by 1
    \xdef#1{\the\referencecount }}
\def\Refnum#1{\REFNUM #1\refend@ } 
\def\REF#1{\REFNUM #1\R@FWRITE\ignorespaces}
\def\Ref#1{\Refnum #1\REFWRITE }
\def\ref{\Ref\?}
\def\REFS#1{\REFNUM #1\gl@bal\lastrefsbegincount=\referencecount
    \REFWRITE }

\def\r@fitem#1{\par \hangafter=0 \hangindent=\refindent \Textindent{#1}}
\def\refitem#1{\r@fitem{#1.}}
\def\NPrefitem#1{\r@fitem{[#1]}}
\def\NPrefs{\let\refmark=\NPrefmark \let\refitem=NPrefitem}
\def\REFWRITE{\R@FWRITE\rel@x }
\def\R@FWRITE#1{\ifreferenceopen \else \gl@bal\referenceopentrue
     \immediate\openout\referencewrite=\jobname.refs
     \toks@={\begingroup \refoutspecials \catcode`\^^M=10 }%
     \immediate\write\referencewrite{\the\toks@}\fi
    \immediate\write\referencewrite{\noexpand\refitem %
                                    {\the\referencecount}}%
    \p@rse@ndwrite \referencewrite #1}
\begingroup
 \catcode`\^^M=\active \let^^M=\relax %
 \gdef\p@rse@ndwrite#1#2{\begingroup \catcode`\^^M=12 \newlinechar=`\^^M%
         \chardef\rw@write=#1\sc@nlines#2}%
 \gdef\sc@nlines#1#2{\sc@n@line \g@rbage #2^^M\endsc@n \endgroup #1}%
 \gdef\sc@n@line#1^^M{\expandafter\toks@\expandafter{\deg@rbage #1}%
         \immediate\write\rw@write{\the\toks@}%
         \futurelet\n@xt \sc@ntest }%
\endgroup
\def\sc@ntest{\ifx\n@xt\endsc@n \let\n@xt=\rel@x
       \else \let\n@xt=\sc@n@notherline \fi \n@xt }
\def\sc@n@notherline{\sc@n@line \g@rbage }
\def\deg@rbage#1{}
\let\g@rbage=\relax    \let\endsc@n=\relax
\def\refout{\par\penalty-400\vskip\chapterskip
   \spacecheck\referenceminspace
   \ifreferenceopen \Closeout\referencewrite \referenceopenfalse \fi
   \line{\fourteenrm\hfil REFERENCES\hfil}\vskip\headskip
   \input \jobname.refs
   }
\def\refoutspecials{\sfcode`\.=1000 \interlinepenalty=1000
         \rightskip=\z@ plus 1em minus \z@ }
\def\Closeout#1{\toks0={\par\endgroup}\immediate\write#1{\the\toks0}%
   \immediate\closeout#1}
%
%
\newcount\figurecount     \figurecount=0
\newcount\tablecount      \tablecount=0
\newif\iffigureopen       \newwrite\figurewrite
\newif\iftableopen        \newwrite\tablewrite
\def\FIGNUM#1{\rel@x \gl@bal\advance\figurecount by 1
    \xdef#1{\the\figurecount}}
\def\FIGURE#1{\FIGNUM #1\F@GWRITE\ignorespaces }

\def\figitem#1{\r@fitem{#1)}}
\def\FIGWRITE{\F@GWRITE\rel@x }
\def\TABNUM#1{\rel@x \gl@bal\advance\tablecount by 1
    \xdef#1{\the\tablecount}}
\def\TABLE#1{\TABNUM #1\T@BWRITE\ignorespaces }
\def\Table{\TABNUM\?Table~\?\TABWRITE }
\def\tabitem#1{\r@fitem{#1:}}
\def\TABWRITE{\T@BWRITE\rel@x }
\def\F@GWRITE#1{\iffigureopen \else \gl@bal\figureopentrue
     \immediate\openout\figurewrite=\jobname.figs
     \toks@={\begingroup \catcode`\^^M=10 }%
     \immediate\write\figurewrite{\the\toks@}\fi
    \immediate\write\figurewrite{\noexpand\figitem %
                                 {\the\figurecount}}%
    \p@rse@ndwrite \figurewrite #1}
\def\T@BWRITE#1{\iftableopen \else \gl@bal\tableopentrue
     \immediate\openout\tablewrite=\jobname.tabs
     \toks@={\begingroup \catcode`\^^M=10 }%
     \immediate\write\tablewrite{\the\toks@}\fi
    \immediate\write\tablewrite{\noexpand\tabitem %
                                 {\the\tablecount}}%
    \p@rse@ndwrite \tablewrite #1}
\def\figout{\par\penalty-400
   \vskip\chapterskip\spacecheck\referenceminspace
   \iffigureopen \Closeout\figurewrite \figureopenfalse \fi
   \line{\fourteenrm\hfil FIGURE CAPTIONS\hfil}\vskip\headskip
   \input \jobname.figs
   }
\def\tabout{\par\penalty-400
   \vskip\chapterskip\spacecheck\referenceminspace
   \iftableopen \Closeout\tablewrite \tableopenfalse \fi
   \line{\fourteenrm\hfil TABLE CAPTIONS\hfil}\vskip\headskip
   \input \jobname.tabs
   }
%
%
%
\newbox\picturebox
\def\p@cht{\ht\picturebox }
\def\p@cwd{\wd\picturebox }
\def\p@cdp{\dp\picturebox }
\newdimen\xshift
\newdimen\yshift
\newdimen\captionwidth
\newskip\captionskip
\captionskip=15pt plus 5pt minus 3pt
\def\fullwidth{\captionwidth=\hsize }
\newtoks\Caption
\newif\ifcaptioned
\newif\ifselfcaptioned
\def\caption{\captionedtrue \Caption }
\newcount\linesabove
\newif\iffileexists
\newtoks\picfilename
\def\fil@#1 {\fileexiststrue \picfilename={#1}}
\def\file#1{\if=#1\let\n@xt=\fil@ \else \def\n@xt{\fil@ #1}\fi \n@xt }
\def\pl@t{\begingroup \pr@tect
    \setbox\picturebox=\hbox{}\fileexistsfalse
    \let\height=\p@cht \let\width=\p@cwd \let\depth=\p@cdp
    \xshift=\z@ \yshift=\z@ \captionwidth=\z@
    \Caption={}\captionedfalse
    \linesabove =0 \picturedefault }
\def\plot{\pl@t \selfcaptionedfalse }
\def\Picture#1{\gl@bal\advance\figurecount by 1
    \xdef#1{\the\figurecount}\pl@t \selfcaptionedtrue }

\def\s@vepicture{\iffileexists \parsefilename \redopicturebox \fi
   \ifdim\captionwidth>\z@ \else \captionwidth=\p@cwd \fi
   \xdef\lastpicture{%
      \iffileexists%
         \setbox0=\hbox{\raise\the\yshift \vbox{%
              \moveright\the\xshift\hbox{\picturedefinition}}}%
      \else%
         \setbox0=\hbox{}%
      \fi
      \ht0=\the\p@cht \wd0=\the\p@cwd \dp0=\the\p@cdp
      \vbox{\hsize=\the\captionwidth%
            \line{\hss\box0 \hss }%
            \ifcaptioned%
               \vskip\the\captionskip \noexpand\Tenpoint
               \ifselfcaptioned%
                   Figure~\the\figurecount.\enspace%
               \fi%
               \the\Caption%
           \fi%
           }%
      }%
      \endgroup%
}
\let\endpicture=\s@vepicture
\def\savepicture#1{\s@vepicture \global\let#1=\lastpicture }
\def\displaypicture{\fullwidth \s@vepicture $$\lastpicture $${}}
\def\toppicture{\fullwidth \s@vepicture \topinsert
    \lastpicture \medskip \endinsert }
\def\midpicture{\fullwidth \s@vepicture \midinsert
    \lastpicture \endinsert }
%
%
\def\leftpicture{\pres@tpicture
    \dimen@i=\hsize \advance\dimen@i by -\dimen@ii
    \setbox\picturebox=\hbox to \hsize {\box0 \hss }%
    \wr@paround }
\def\rightpicture{\pres@tpicture
    \dimen@i=\z@
    \setbox\picturebox=\hbox to \hsize {\hss \box0 }%
    \wr@paround }
\def\pres@tpicture{\gl@bal\linesabove=\linesabove
    \s@vepicture \setbox\picturebox=\vbox{
         \kern \linesabove\baselineskip \kern 0.3\baselineskip
         \lastpicture \kern 0.3\baselineskip }%
    \dimen@=\p@cht \dimen@i=\dimen@
    \advance\dimen@i by \pagetotal
    \par \ifdim\dimen@i>\pagegoal \vfil\break \fi
    \dimen@ii=\hsize
    \advance\dimen@ii by -\parindent \advance\dimen@ii by -\p@cwd
    \setbox0=\vbox to\z@{\kern-\baselineskip \unvbox\picturebox \vss }}
\def\wr@paround{\Caption={}\count255=1
    \loop \ifnum \linesabove >0
         \advance\linesabove by -1 \advance\count255 by 1
         \advance\dimen@ by -\baselineskip
         \expandafter\Caption \expandafter{\the\Caption \z@ \hsize }%
      \repeat
    \loop \ifdim \dimen@ >\z@
         \advance\count255 by 1 \advance\dimen@ by -\baselineskip
         \expandafter\Caption \expandafter{%
             \the\Caption \dimen@i \dimen@ii }%
      \repeat
    \edef\n@xt{\parshape=\the\count255 \the\Caption \z@ \hsize }%
    \par\noindent \n@xt \strut \vadjust{\box\picturebox }}
\let\picturedefault=\relax
\let\parsefilename=\relax
\def\redopicturebox{\let\picturedefinition=\rel@x
   \errhelp=\disabledpictures
   \errmessage{This version of TeX cannot handle pictures.  Sorry.}}
\newhelp\disabledpictures
     {You will get a blank box in place of your picture.}
%
%
%
%
%
%
%
%
%
%
\def\FRONTPAGE{\ifvoid255\else\vfill\penalty-20000\fi
   \gl@bal\pagenumber=1     \gl@bal\chapternumber=0
   \gl@bal\equanumber=0     \gl@bal\sectionnumber=0
   \gl@bal\referencecount=0 \gl@bal\figurecount=0
   \gl@bal\tablecount=0     \gl@bal\frontpagetrue
   \gl@bal\lastf@@t=0       \gl@bal\footsymbolcount=0
   \gl@bal\cn@@false }

\def\papers{\papersize\headline=\paperheadline\footline=\paperfootline}
\def\papersize{\hsize=35pc \vsize=50pc \hoffset=0pc \voffset=1pc
   \advance\hoffset by\HOFFSET \advance\voffset by\VOFFSET
   \pagebottomfiller=0pc
   \skip\footins=\bigskipamount \normalspace }
\papers  
%
%
\newskip\lettertopskip       \lettertopskip=20pt plus 50pt
\newskip\letterbottomskip    \letterbottomskip=\z@ plus 100pt
\newskip\signatureskip       \signatureskip=40pt plus 3pt
\def\lettersize{\hsize=6.5in \vsize=8.5in \hoffset=0in \voffset=0.5in
   \advance\hoffset by\HOFFSET \advance\voffset by\VOFFSET
   \pagebottomfiller=\letterbottomskip
   \skip\footins=\smallskipamount \multiply\skip\footins by 3
   \singlespace }
\def\MEMO{\lettersize \headline=\letterheadline \footline={\hfil }%
   \let\rule=\memorule \FRONTPAGE \memohead }

\def\memodate{\afterassignment\MEMO \date }
\def\memit@m#1{\smallskip \hangafter=0 \hangindent=1in
    \Textindent{\caps #1}}
\def\subject{\memit@m{Subject:}}
\def\topic{\memit@m{Topic:}}
\def\from{\memit@m{From:}}
\def\to{\rel@x \ifmmode \rightarrow \else \memit@m{To:}\fi }
\def\memorule{\medskip\hrule height 1pt\bigskip}  
\def\memohead{\centerline{\fourteenrm MEMORANDUM}}
\newwrite\labelswrite
\newtoks\rw@toks
\def\letters{\lettersize
   \headline=\letterheadline \footline=\letterfootline
   \immediate\openout\labelswrite=\jobname.lab}

\let\letterhead=\rel@x
\def\addressee#1{\medskip\line{\hskip 0.75\hsize plus\z@ minus 0.25\hsize
                               \the\date \hfil }%
   \vskip \lettertopskip
   \ialign to\hsize{\strut ##\hfil\tabskip 0pt plus \hsize \crcr #1\crcr}
   \writelabel{#1}\medskip \noindent\hskip -\spaceskip \ignorespaces }
\def\rwl@begin#1\cr{\rw@toks={#1\crcr}\rel@x
   \immediate\write\labelswrite{\the\rw@toks}\futurelet\n@xt\rwl@next}
\def\rwl@next{\ifx\n@xt\rwl@end \let\n@xt=\rel@x
      \else \let\n@xt=\rwl@begin \fi \n@xt}
\let\rwl@end=\rel@x
\def\writelabel#1{\immediate\write\labelswrite{\noexpand\labelbegin}
     \rwl@begin #1\cr\rwl@end
     \immediate\write\labelswrite{\noexpand\labelend}}
\newtoks\FromAddress         \FromAddress={}
\newtoks\sendername          \sendername={}
\newbox\FromLabelBox
\newdimen\labelwidth          \labelwidth=6in
\def\makelabels{\afterassignment\Makelabels \sendername=}
\def\Makelabels{\FRONTPAGE \letterinfo={\hfil } \MakeFromBox
     \immediate\closeout\labelswrite  \input \jobname.lab\vfil\eject}
\let\labelend=\rel@x
\def\labelbegin#1\labelend{\setbox0=\vbox{\ialign{##\hfil\cr #1\crcr}}
     \MakeALabel }
\def\MakeFromBox{\gl@bal\setbox\FromLabelBox=\vbox{\Tenpoint
     \ialign{##\hfil\cr \the\sendername \the\FromAddress \crcr }}}
\def\MakeALabel{\vskip 1pt \hbox{\vrule \vbox{
        \hsize=\labelwidth \hrule\bigskip
        \leftline{\hskip 1\parindent \copy\FromLabelBox}\bigskip
        \centerline{\hfil \box0 } \bigskip \hrule
        }\vrule } \vskip 1pt plus 1fil }
\def\signed#1{\par \nobreak \bigskip \dt@pfalse \begingroup
  \everycr={\noalign{\nobreak
            \ifdt@p\vskip\signatureskip\gl@bal\dt@pfalse\fi }}%
  \tabskip=0.5\hsize plus \z@ minus 0.5\hsize
  \halign to\hsize {\strut ##\hfil\tabskip=\z@ plus 1fil minus \z@\crcr
          \noalign{\gl@bal\dt@ptrue}#1\crcr }%
  \endgroup \bigskip }
\newbox\letterb@x
\def\lettertext{\par \vskip\parskip \unvcopy\letterb@x \par }
\def\multiletter{\setbox\letterb@x=\vbox\bgroup
      \everypar{\vrule height 1\baselineskip depth 0pt width 0pt }
      \singlespace \topskip=\baselineskip }
\def\letterend{\par\egroup}
%
%
%
\newskip\frontpageskip
\newtoks\Pubnum   
\newtoks\Pubtype  \let\pubtype=\Pubtype
\newif\ifp@bblock  \p@bblocktrue
\def\PH@SR@V{\doubl@true \baselineskip=24.1pt plus 0.2pt minus 0.1pt
             \parskip= 3pt plus 2pt minus 1pt }
\def\PHYSREV{\papers\PhysRevtrue\PH@SR@V}

\def\titlepage{\FRONTPAGE\papers\ifPhysRev\PH@SR@V\fi
   \ifp@bblock\p@bblock \else\hrule height\z@ \rel@x \fi }
\def\nopubblock{\p@bblockfalse}
\def\endpage{\vfil\break}
\frontpageskip=12pt plus .5fil minus 2pt
\Pubtype={}
\Pubnum={}
\def\p@bblock{\begingroup \tabskip=\hsize minus \hsize
   \baselineskip=1.5\ht\strutbox \topspace-2\baselineskip
   \halign to\hsize{\strut ##\hfil\tabskip=0pt\crcr
       \the\Pubnum\crcr\the\date\crcr\the\pubtype\crcr}\endgroup}
\def\title#1{\vskip\frontpageskip \titlestyle{#1} \vskip\headskip }
\def\author#1{\vskip\frontpageskip\titlestyle{\twelvecp #1}\nobreak}

\def\address#1{\par\kern 5pt\titlestyle{\twelvepoint\it #1}}
\def\andaddress{\par\kern 5pt \centerline{\sl and} \address}

\def\abstract{\par\dimen@=\prevdepth \hrule height\z@ \prevdepth=\dimen@
   \vskip\frontpageskip\centerline{\fourteenrm ABSTRACT}\vskip\headskip }

%
%
%
\def\ie{\hbox{\it i.e.}}       
\def\eg{\hbox{\it e.g.}}       
   
\def\\{\rel@x \ifmmode \backslash \else {\tt\char`\\}\fi }
\def\sequentialequations{\rel@x \if\equanumber<0 \else
  \gl@bal\equanumber=-\equanumber \gl@bal\advance\equanumber by -1 \fi }
\def\journal#1&#2(#3){\begingroup \let\journal=\dummyj@urnal
    \unskip, \sl #1\unskip~\bf\ignorespaces #2\rm
    (\afterassignment\j@ur \count255=#3), \endgroup\ignorespaces }
\def\j@ur{\ifnum\count255<100 \advance\count255 by 1900 \fi
          \number\count255 }
\def\dummyj@urnal{%
    \toks@={Reference foul up: nested \journal macros}%
    \errhelp={Your forgot & or ( ) after the last \journal}%
    \errmessage{\the\toks@ }}

\def\topspace{\hrule height 0pt depth 0pt \vskip}

\def\Buildrel#1\under#2{\mathrel{\mathop{#2}\limits_{#1}}}
\def\becomes#1{\mathchoice{\becomes@\scriptstyle{#1}}
   {\becomes@\scriptstyle{#1}} {\becomes@\scriptscriptstyle{#1}}
   {\becomes@\scriptscriptstyle{#1}}}
\def\becomes@#1#2{\mathrel{\setbox0=\hbox{$\m@th #1{\,#2\,}$}%
        \mathop{\hbox to \wd0 {\rightarrowfill}}\limits_{#2}}}

\def\VEV#1{\left\langle #1\right\rangle}

\let\int=\intop         
\def\lsim{\mathrel{\mathpalette\@versim<}}
\def\gsim{\mathrel{\mathpalette\@versim>}}
\def\@versim#1#2{\vcenter{\offinterlineskip
        \ialign{$\m@th#1\hfil##\hfil$\crcr#2\crcr\sim\crcr } }}
\def\big#1{{\hbox{$\left#1\vbox to 0.85\b@gheight{}\right.\n@space$}}}
\def\Big#1{{\hbox{$\left#1\vbox to 1.15\b@gheight{}\right.\n@space$}}}
\def\bigg#1{{\hbox{$\left#1\vbox to 1.45\b@gheight{}\right.\n@space$}}}
\def\Bigg#1{{\hbox{$\left#1\vbox to 1.75\b@gheight{}\right.\n@space$}}}
\def\){\mskip 2mu\nobreak }
%
%
%
\let\sec@nt=\sec
\def\sec{\rel@x\ifmmode\let\n@xt=\sec@nt\else\let\n@xt\section\fi\n@xt}
\def\obsolete#1{\message{Macro \string #1 is obsolete.}}
\def\firstsec#1{\obsolete\firstsec \section{#1}}
\def\firstsubsec#1{\obsolete\firstsubsec \subsection{#1}}
\def\thispage#1{\obsolete\thispage \gl@bal\pagenumber=#1\frontpagefalse}
\def\thischapter#1{\obsolete\thischapter \gl@bal\chapternumber=#1}
\def\splitout{\obsolete\splitout\rel@x}
\def\prop{\obsolete\prop \propto }
\def\nextequation#1{\obsolete\nextequation \gl@bal\equanumber=#1
   \ifnum\the\equanumber>0 \gl@bal\advance\equanumber by 1 \fi}
\def\BOXITEM{\afterassigment\B@XITEM\setbox0=}
\def\B@XITEM{\par\hangindent\wd0 \noindent\box0 }
%
%
%
\def\phyzzx{PHY\setbox0=\hbox{Z}\copy0 \kern-0.5\wd0 \box0 X}
        
\everyjob{\xdef\today{\monthname~\number\day, \number\year}
        \input myphyx.tex }
\message{ by V.K.}
%
%
%
   \message{V 1.18 mods and bug fixes by M.Weinstein}
   \def\unlock{\catcode`@=11}

   \def\lock{\catcode`@=12}

   \unlock
%
%
   \def\PRrefmark#1{~[#1]}
   \def\refitem#1{\ifPhysRev\r@fitem{[#1]}\else\r@fitem{#1.}\fi}
   \def\generatefootsymbol{%
      \ifcase\footsymbolcount%
          \fd@f 13F \or \fd@f 279 \or \fd@f 27A %
            \or \fd@f 278 \or \fd@f 27B %
      \else
         \ifnum\footsymbolcount <0 %
            \xdef\footsymbol{\number-\footsymbolcount}
         \else %
            \fd@f 203
               {\loop \ifnum\footsymbolcount >5
                  \fd@f{203 \footsymbol } %
                  \advance\footsymbolcount by -1%
                \repeat %
               }%
         \fi%
      \fi%
   }
   \def\OldPhysRevRefmark{\let\PRrefmark=\attach}
   \def\OldPRRefitem#1{\r@fitem{#1.}}
   \def\OldPhysRevRefitem{\let\refitem=\OldPRRefitem}
   \def\NPrefs{\let\refmark=\NPrefmark \let\refitem=\NPrefitem}
%
    \newif\iffileexists              \fileexistsfalse
    \newif\ifforwardrefson           \forwardrefsontrue
    \newif\ifamiga                   \amigatrue
    \newif\iflinkedinput             \linkedinputtrue
    \newif\iflinkopen                \linkopenfalse
    \newif\ifcsnameopen              \csnameopenfalse
    \newif\ifdummypictures           \dummypicturesfalse
    \newif\ifcontentson              \contentsonfalse
    \newif\ifcontentsopen            \contentsopenfalse
    \newif\ifmakename                \makenamefalse
    \newif\ifverbdone
    \newif\ifusechapterlabel         \usechapterlabelfalse
    \newif\ifstartofchapter          \startofchapterfalse
    \newif\iftableofplates           \tableofplatesfalse
    \newif\ifplatesopen              \platesopenfalse
    \newif\iftableoftables           \tableoftablesfalse
    \newif\iftableoftablesopen       \tableoftablesopenfalse
    \newif\ifwarncsname              \warncsnamefalse
%
    \newwrite\linkwrite
    \newwrite\csnamewrite
    \newwrite\contentswrite
    \newwrite\plateswrite
    \newwrite\tableoftableswrite
    \newread\testifexists
    \newread\verbinfile

    \newtoks\jobdir                  \jobdir={}
    \newtoks\tempnametoks            \tempnametoks={}
    \newtoks\oldheadline             \oldheadline={}
    \newtoks\oldfootline             \oldfootline={}
    \newtoks\subsectstyle            \subsectstyle={\Number}
    \newtoks\subsubsectstyle         \subsubsectstyle={\Number}
    \newtoks\runningheadlines        \runningheadlines={\relax}
    \newtoks\chapterformat           \chapterformat={\titlestyle}
    \newtoks\sectionformat           \sectionformat={\relax}
    \newtoks\subsectionformat        \subsectionformat={\relax}
    \newtoks\subsubsectionformat     \subsubsectionformat={\relax}
    \newtoks\chapterfontstyle        \chapterfontstyle={\bf}
    \newtoks\sectionfontstyle        \sectionfontstyle={\rm}
    \newtoks\subsectionfontstyle     \subsectionfontstyle={\rm}
    \newtoks\sectionfontstyleb       \sectionfontstyleb={\caps}
    \newtoks\subsubsectionfontstyle  \subsubsectionfontstyle={\rm}

    \newcount\subsectnumber           \subsectnumber=0
    \newcount\subsubsectnumber        \subsubsectnumber=0


   \newdimen\pictureindent           \pictureindent=15pt
   \newdimen\str
   \newdimen\squareht
   \newdimen\squarewd
   \newskip\doublecolskip
   \newskip\tableoftablesskip        \tableoftablesskip=\baselineskip


   \newbox\squarebox


   \newskip\sectionindent            \sectionindent=0pt
   \newskip\subsectionindent         \subsectionindent=0pt
  \def\thechapterhead{\relax}
  \def\thesectionhead{\relax}
  \def\thesubsecthead{\relax}
  \def\thesubsubsecthead{\relax}


   \def\GetIfExists #1 {
       \immediate\openin\testifexists=#1
       \ifeof\testifexists
           \immediate\closein\testifexists
       \else
         \immediate\closein\testifexists
         \input #1
       \fi
   }


   \def\stripbackslash#1#2*{\def\strippedname{#2}}

   \def\ifundefined#1{\expandafter\ifx\csname#1\endcsname\relax}

   \def\val#1{%
      \expandafter\stripbackslash\string#1*%
      \ifundefined{\strippedname}%
      \message{Warning! The control sequence \noexpand#1 is not defined.} ? %
      \else\csname\strippedname\endcsname\fi%
   }
%
%
   \def\CheckForOverWrite#1{%
      \expandafter\stripbackslash\string#1*%
      \ifundefined{\strippedname}%
      \else%
         \ifwarncsname
            \message{Warning! The control sequence \noexpand#1 is being
          overwritten.}%
          \else
          \fi
      \fi%
   }

   \def\FootNoteFonts{\Tenpoint}

   \def\Vfootnote#1{%
      \insert\footins%
      \bgroup%
         \interlinepenalty=\interfootnotelinepenalty%
         \floatingpenalty=20000%
         \singl@true\doubl@false%
         \FootNoteFonts%
         \splittopskip=\ht\strutbox%
         \boxmaxdepth=\dp\strutbox%
         \leftskip=\footindent%
         \rightskip=\z@skip%
         \parindent=0.5%
         \footindent%
         \parfillskip=0pt plus 1fil%
         \spaceskip=\z@skip%
         \xspaceskip=\z@skip%
         \footnotespecial%
         \Textindent{#1}%
         \footstrut%
         \futurelet\next\fo@t%
   }

   \def\csnamech@ck{%
       \ifcsnameopen%
       \else%
           \global\csnameopentrue%
           \immediate\openout\csnamewrite=\the\jobdir\jobname.csnames%
           \immediate\write\csnamewrite{\unlock}%
       \fi%
   }

   \def\linksch@ck{%
          \iflinkopen%
          \else%
              \global\linkopentrue%
              \immediate\openout\linkwrite=\the\jobdir\jobname.links%
          \fi%
   }

   \def\c@ntentscheck{%
       \ifcontentsopen%
       \else%
           \global\contentsopentrue%
           \immediate\openout\contentswrite=\the\jobdir\jobname.contents%
           \immediate\write\contentswrite{%
                \noexpand\titlestyle{Table of Contents}%
           }%
           \immediate\write\contentswrite{\noexpand\bigskip}%
       \fi%
   }

   \def\t@bleofplatescheck{%
       \ifplatesopen%
       \else%
           \global\platesopentrue%
           \immediate\openout\plateswrite=\the\jobdir\jobname.plates%
           \immediate\write\plateswrite{%
                \noexpand\titlestyle{Illustrations}%
           }%
           \immediate\write\plateswrite{%
              \unlock%
           }%
           \immediate\write\plateswrite{\noexpand\bigskip}%
       \fi%
   }

   \def\t@bleoftablescheck{%
       \iftableoftablesopen%
       \else%
           \global\tableoftablesopentrue%
          \immediate\openout\tableoftableswrite=\the\jobdir\jobname.tables%
           \immediate\write\tableoftableswrite{%
                \noexpand\titlestyle{Tables}%
           }%
           \immediate\write\tableoftableswrite{%
              \unlock%
           }%
           \immediate\write\tableoftableswrite{\noexpand\bigskip}%
       \fi%
   }


   \def\linkinput#1 {\input #1
       \iflinkedinput \relax \else \global\linkedinputtrue \fi
       \linksch@ck
       \immediate\write\linkwrite{#1}
   }


   \def\fil@#1 {%
       \ifdummypictures%
          \fileexistsfalse%
          \picfilename={}%
       \else%
          \fileexiststrue%
          \picfilename={#1}%
       \fi%
       \iflinkedinput%
          \iflinkopen \relax%
          \else%
            \global\linkopentrue%
            \immediate\openout\linkwrite=\the\jobdir\jobname.links%
          \fi%
          \immediate\write\linkwrite{#1}%
       \fi%
   }
   \def\Picture#1{%
      \gl@bal\advance\figurecount by 1%
      \CheckForOverWrite#1%
      \xdef#1{\the\figurecount}\pl@t%
      \selfcaptionedtrue%
   }

   \def\s@vepicture{%
       \iffileexists \parsefilename \redopicturebox \fi%
       \ifdim\captionwidth>\z@ \else \captionwidth=\p@cwd \fi%
       \xdef\lastpicture{%
          \iffileexists%
             \setbox0=\hbox{\raise\the\yshift \vbox{%
                \moveright\the\xshift\hbox{\picturedefinition}}%
             }%
          \else%
             \setbox0=\hbox{}%
          \fi
          \ht0=\the\p@cht \wd0=\the\p@cwd \dp0=\the\p@cdp%
          \vbox{\hsize=\the\captionwidth \line{\hss\box0 \hss }%
          \ifcaptioned%
             \vskip\the\captionskip \noexpand\Tenpoint%
             \ifselfcaptioned%
                Figure~\the\figurecount.\enspace%
             \fi%
             \the\Caption%
          \fi }%
       }%
       \iftableofplates%
          \ifplatesopen%
          \else%
             \t@bleofplatescheck%
          \fi%
          \ifselfcaptioned%
             \immediate\write\plateswrite{%
                \noexpand\platetext{%
                \noexpand\item{\rm \the\figurecount .}%
                \the\Caption}{\the\pageno}%
             }%
          \else%
             \immediate\write\plateswrite{%
                \noexpand\platetext{\the\Caption}{\the\pageno}%
             }%
          \fi%
       \fi%
       \endgroup%
   }

   \def\platesout{%
      \ifplatesopen
         \immediate\closeout\plateswrite%
         \global\platesopenfalse%
      \fi%
      \input \jobname.plates%
      \lock%
   }

   \def\platetext#1#2{%
       \hbox to \hsize{\vbox{\hsize=.9\hsize #1}\hfill#2}%
       \vskip \tableoftablesskip \vskip\parskip%
   }


   \def\pres@tpicture{%
       \gl@bal\linesabove=\linesabove
       \s@vepicture
       \setbox\picturebox=\vbox{
       \kern \linesabove\baselineskip \kern 0.3\baselineskip
       \lastpicture \kern 0.3\baselineskip }%
       \dimen@=\p@cht \dimen@i=\dimen@
       \advance\dimen@i by \pagetotal
       \par \ifdim\dimen@i>\pagegoal \vfil\break \fi
       \dimen@ii=\hsize
       \advance\dimen@ii by -\pictureindent \advance\dimen@ii by -\p@cwd
       \setbox0=\vbox to\z@{\kern-\baselineskip \unvbox\picturebox \vss }
   }

   \def\subspaces@t#1:#2;{%
      \baselineskip = \normalbaselineskip%
      \multiply\baselineskip by #1 \divide\baselineskip by #2%
      \lineskip = \normallineskip%
      \multiply\lineskip by #1 \divide\lineskip by #2%
      \lineskiplimit = \normallineskiplimit%
      \multiply\lineskiplimit by #1 \divide\lineskiplimit by #2%
      \parskip = \normalparskip%
      \multiply\parskip by #1 \divide\parskip by #2%
      \abovedisplayskip = \normaldisplayskip%
      \multiply\abovedisplayskip by #1 \divide\abovedisplayskip by #2%
      \belowdisplayskip = \abovedisplayskip%
      \abovedisplayshortskip = \normaldispshortskip%
      \multiply\abovedisplayshortskip by #1%
        \divide\abovedisplayshortskip by #2%
      \belowdisplayshortskip = \abovedisplayshortskip%
      \advance\belowdisplayshortskip by \belowdisplayskip%
      \divide\belowdisplayshortskip by 2%
      \smallskipamount = \skipregister%
      \multiply\smallskipamount by #1 \divide\smallskipamount by #2%
      \medskipamount = \smallskipamount \multiply\medskipamount by 2%
      \bigskipamount = \smallskipamount \multiply\bigskipamount by 4%
   }


   \def\makename#1{
       \global\makenametrue
       \global\tempnametoks={#1}
   }

   \def\nomakename#1{\relax}


   \def\savename#1{%
      \CheckForOverWrite{#1}%
      \csnamech@ck%
      \immediate\write\csnamewrite{\def\the\tempnametoks{#1}}%
   }

   \def\FootNoteFonts{\Tenpoint}

   \def\Vfootnote#1{%
      \insert\footins%
      \bgroup%
         \interlinepenalty=\interfootnotelinepenalty%
         \floatingpenalty=20000%
         \singl@true\doubl@false%
         \FootNoteFonts%
         \splittopskip=\ht\strutbox%
         \boxmaxdepth=\dp\strutbox%
         \leftskip=\footindent%
         \rightskip=\z@skip%
         \parindent=0.5%
         \footindent%
         \parfillskip=0pt plus 1fil%
         \spaceskip=\z@skip%
         \xspaceskip=\z@skip%
         \footnotespecial%
         \Textindent{#1}%
         \footstrut%
         \futurelet\next\fo@t%
   }
%

   \def\eqname#1{%
      \CheckForOverWrite{#1}%
      \rel@x{\pr@tect%
      \csnamech@ck%
      \ifnum\equanumber<0%
          \xdef#1{{\noexpand\f@m0(\number-\equanumber)}}%
          \immediate\write\csnamewrite{%
            \def\noexpand#1{\noexpand\f@m0 (\number-\equanumber)}}%
          \gl@bal\advance\equanumber by -1%
      \else%
          \gl@bal\advance\equanumber by 1%
          \ifusechapterlabel%
            \xdef#1{{\noexpand\f@m0(\ifcn@@ \chapterlabel.\fi%
               \number\equanumber)}%
            }%
          \else%
             \xdef#1{{\noexpand\f@m0(\ifcn@@%
                 {\the\chapterstyle{\the\chapternumber}}.\fi%
                 \number\equanumber)}}%
          \fi%
          \ifcn@@%
             \ifusechapterlabel
                \immediate\write\csnamewrite{\def\noexpand#1{(%
                  {\chapterlabel}.%
                  \number\equanumber)}%
                }%
             \else
                \immediate\write\csnamewrite{\def\noexpand#1{(%
                  {\the\chapterstyle{\the\chapternumber}}.%
                  \number\equanumber)}%
                }%
             \fi%
          \else%
              \immediate\write\csnamewrite{\def\noexpand#1{(%
                  \number\equanumber)}}%
          \fi%
      \fi}%
      #1%
   }

   \def\eqn{\eqno\eqname}

   \let\eqnalign=\eqname


   \def\APPENDIX#1#2{%
       \global\usechapterlabeltrue%
       \par\penalty-300\vskip\chapterskip%
       \spacecheck\chapterminspace%
       \chapterreset%
       \xdef\chapterlabel{#1}%
       \titlestyle{APPENDIX #2}%
       \nobreak\vskip\headskip \penalty 30000%
       \wlog{\string\Appendix~\chapterlabel}%
   }

   \def\REFNUM#1{%
      \CheckForOverWrite{#1} %
      \rel@x\gl@bal\advance\referencecount by 1%
      \xdef#1{\the\referencecount}%
      \csnamech@ck%
      \immediate\write\csnamewrite{\def\noexpand#1{\the\referencecount}}%
   }

   %

   \def\FIGNUM#1{
      \CheckForOverWrite{#1}%
      \rel@x\gl@bal\advance\figurecount by 1%
      \xdef#1{\the\figurecount}%
      \csnamech@ck%
      \immediate\write\csnamewrite{\def\noexpand#1{\the\figurecount}}%
   }


   \def\TABNUM#1{%
      \CheckForOverWrite{#1}%
      \rel@x \gl@bal\advance\tablecount by 1%
      \xdef#1{\the\tablecount}%
      \csnamech@ck%
      \immediate\write\csnamewrite{\def\noexpand#1{\the\tablecount}}%
   }


   \def\tableoftableson{%
      \global\tableoftablestrue%

      \gdef\TABLE##1##2{%
         \t@bleoftablescheck%
         \TABNUM ##1%
         \immediate\write\tableoftableswrite{%
            \noexpand\tableoftablestext{%
            \noexpand\item{\rm \the\tablecount .}%
                ##2}{\the\pageno}%
             }%
      }

      \gdef\Table##1{\TABLE\?{##1}Table~\?}
   }

   \def\tableoftablestext#1#2{%
       \hbox to \hsize{\vbox{\hsize=.9\hsize #1}\hfill#2}%
       \vskip \tableoftablesskip%
   }

   \def\tableoftablesout{%
      \iftableoftablesopen
         \immediate\closeout\tableoftableswrite%
         \global\tableoftablesopenfalse%
      \fi%
      \input \jobname.tables%
      \lock%
   }

%
%
%
%
%
%

   \def\contentsoff{\contentsonfalse}

   \def\f@m#1{\f@ntkey=#1\fam=\f@ntkey\the\textfont\f@ntkey\rel@x}
   \def\em@{\rel@x%
      \ifnum\f@ntkey=0\it%
      \else%
         \ifnum\f@ntkey=\bffam\it%
         \else\rm  %
         \fi%
      \fi%
   }

   \def\fontsoff{%
      \def\mit{\relax}%
      \let\oldstyle=\mit%
      \def\cal{\relax}%
      \def\it{\relax}%
      \def\sl{\relax}%
      \def\bf{\relax}%
      \def\tt{\relax}%
      \def\caps{\relax}%
      \let\cp=\caps%
   }


   \def\fontson{%
      \def\rm{\n@expand\f@m0}%
      \def\mit{\n@expand\f@m1}%
      \let\oldstyle=\mit%
      \def\cal{\n@expand\f@m2}%
      \def\it{\n@expand\f@m\itfam}%
      \def\sl{\n@expand\f@m\slfam}%
      \def\bf{\n@expand\f@m\bffam}%
      \def\tt{\n@expand\f@m\ttfam}%
      \def\caps{\n@expand\f@m\cpfam}%
      \let\cp=\caps%
   }

   \fontson
%


   \def\@alpha#1{\count255='140 \advance\count255 by #1\char\count255}
   \def\alphabetic{\@alpha}
   \def\@Alpha#1{\count255='100 \advance\count255 by #1\char\count255}
   \def\Alphabetic{\@Alpha}
   \def\@Roman#1{\uppercase\expandafter{\romannumeral #1}}
   \def\Roman{\@Roman}
   \def\@roman#1{\romannumeral #1}
   \def\roman{\@roman}
   \def\@number#1{\number #1}
   \def\Number{\@number}

   \def\leaderfill{\leaders\hbox to 1em{\hss.\hss}\hfill}

   \def\chapterinfo#1{%
      \line{%
         \ifcn@@%
            \hbox to \itemsize{\hfil\chapterlabel .\quad\ }%
         \fi%
         \noexpand{#1}\leaderfill\the\pagenumber%
      }%
   }

   \def\sectioninfo#1{%
      \line{%
         \ifcn@@%
            \hbox to 2\itemsize{\hfil\sectlabel \quad}%
          \else%
            \hbox to \itemsize{\hfil\quad}%
          \fi%
          \ \noexpand{#1}%
          \leaderfill \the\pagenumber%
      }%
   }

   \def\subsectioninfo#1{%
      \line{%
         \ifcn@@%
            \hbox to 3\itemsize{\hfil \quad\subsectlabel\quad}%
         \else%
            \hbox to 2\itemsize{\hfil\quad}%
         \fi%
          \ \noexpand{#1}%
          \leaderfill \the\pagenumber%
      }%
   }

   \def\subsubsecinfo#1{%
      \line{%
         \ifcn@@%
            \hbox to 4\itemsize{\hfil\subsubsectlabel\quad}%
         \else%
            \hbox to 3\itemsize{\hfil\quad}%
         \fi%
         \ \noexpand{#1}\leaderfill \the\pagenumber%
      }%
   }

   \def\CONTENTS#1;#2{
       {\let\makename=\nomakename
        \if#1C
            \immediate\write\contentswrite{\chapterinfo{#2}}%
        \else\if#1S
                \immediate\write\contentswrite{\sectioninfo{#2}}%
             \else\if#1s
                     \immediate\write\contentswrite{\subsectioninfo{#2}}%
                  \else\if#1x
                          \immediate\write\contentswrite{%
                              \subsubsecinfo{#2}}%
                       \fi
                  \fi
             \fi
        \fi
       }
   }

   \def\chapterreset{\gl@bal\advance\chapternumber by 1%
       \ifnum\equanumber<0 \else\gl@bal\equanumber=0 \fi%
       \gl@bal\sectionnumber=0  \gl@bal\let\sectlabel=\rel@x%
       \gl@bal\subsectnumber=0   \gl@bal\let\subsectlabel=\rel@x%
       \gl@bal\subsubsectnumber=0 \gl@bal\let\subsubsectlabel=\rel@x%
       \ifcn@%
           \gl@bal\cn@@true {\pr@tect\xdef\chapterlabel{%
           {\the\chapterstyle{\the\chapternumber}}}}%
       \else%
           \gl@bal\cn@@false \gdef\chapterlabel{\rel@x}%
       \fi%
       \gl@bal\startofchaptertrue%
   }

   \def\chapter#1{\par \penalty-300 \vskip\chapterskip%
       \spacecheck\chapterminspace%
       \gdef\thechapterhead{#1}%
       \gdef\thesectionhead{\relax}%
       \gdef\thesubsecthead{\relax}%
       \gdef\thesubsubsecthead{\relax}%
       \chapterreset \the\chapterformat{\the\chapterfontstyle%
          \ifcn@@\chapterlabel.~~\fi #1}%
       \nobreak\vskip\headskip \penalty 30000%
       {\pr@tect\wlog{\string\chapter\space \chapterlabel}}%
       \ifmakename%
           \csnamech@ck
           \ifcn@@%
              \immediate\write\csnamewrite{\def\the\tempnametoks{%
                 {\the\chapterstyle{\the\chapternumber}}}%
              }%
            \fi%
            \global\makenamefalse%
       \fi%
       \ifcontentson%
          \c@ntentscheck%
          \CONTENTS{C};{#1}%
       \fi%
       }%

   \def\section#1{\par \ifnum\lastpenalty=30000\else%
       \penalty-200\vskip\sectionskip \spacecheck\sectionminspace\fi%
       \gl@bal\advance\sectionnumber by 1%
       \gl@bal\subsectnumber=0%
       \gl@bal\let\subsectlabel=\rel@x%
       \gl@bal\subsubsectnumber=0%
       \gl@bal\let\subsubsectlabel=\rel@x%
       \gdef\thesectionhead{#1}%
       \gdef\thesubsecthead{\relax}%
       \gdef\thesubsubsecthead{\relax}%
       {\pr@tect\xdef\sectlabel{\ifcn@@%
          {\the\chapterstyle{\the\chapternumber}}.%
          {\the\sectionstyle{\the\sectionnumber}}\fi}%
       \wlog{\string\section\space \sectlabel}}%
       \the\sectionformat{\noindent\the\sectionfontstyle%
            {\ifcn@@\unskip\hskip\sectionindent\sectlabel~~\fi%
                \the\sectionfontstyleb#1}}%
       \par%
       \nobreak\vskip\headskip \penalty 30000%
       \ifmakename%
           \csnamech@ck%
           \ifcn@@%
              \immediate\write\csnamewrite{\def\the\tempnametoks{%
                 {\the\chapterstyle{\the\chapternumber}.%
                  \the\sectionstyle{\the\sectionnumber}}}
              }%
            \fi%
            \global\makenamefalse%
       \fi%
       \ifcontentson%
          \c@ntentscheck%
          \CONTENTS{S};{#1}%
       \fi%
   }

   \def\subsection#1{\par \ifnum\lastpenalty=30000\else%
       \penalty-200\vskip\sectionskip \spacecheck\sectionminspace\fi%
       \gl@bal\advance\subsectnumber by 1%
       \gl@bal\subsubsectnumber=0%
       \gl@bal\let\subsubsectlabel=\rel@x%
       \gdef\thesubsecthead{#1}%
       \gdef\thesubsubsecthead{\relax}%
       {\pr@tect\xdef\subsectlabel{\the\subsectionfontstyle%
           \ifcn@@{\the\chapterstyle{\the\chapternumber}}.%
           {\the\sectionstyle{\the\sectionnumber}}.%
           {\the\subsectstyle{\the\subsectnumber}}\fi}%
           \wlog{\string\section\space \subsectlabel}%
       }%
       \the\subsectionformat{\noindent\the\subsectionfontstyle%
         {\ifcn@@\unskip\hskip\subsectionindent%
          \subsectlabel~~\fi#1}}%
       \par%
       \nobreak\vskip\headskip \penalty 30000%
       \ifmakename%
           \csnamech@ck%
           \ifcn@@%
              \immediate\write\csnamewrite{\def\the\tempnametoks{%
                 {\the\chapterstyle{\the\chapternumber}}.%
                 {\the\sectionstyle{\the\sectionnumber}}.%
                 {\the\subsectstyle{\the\subsectnumber}}}%
              }%
            \fi%
            \global\makenamefalse%
       \fi%
       \ifcontentson%
          \c@ntentscheck%
          \CONTENTS{s};{#1}%
       \fi%
   }

   \def\subsubsection#1{\par \ifnum\lastpenalty=30000\else%
       \penalty-200\vskip\sectionskip \spacecheck\sectionminspace\fi%
       \gl@bal\advance\subsubsectnumber by 1%
       \gdef\thesubsubsecthead{#1}%
       {\pr@tect\xdef\subsubsectlabel{\the\subsubsectionfontstyle\ifcn@@%
           {\the\chapterstyle{\the\chapternumber}}.%
           {\the\sectionstyle{\the\sectionnumber}}.%
           {\the\subsectstyle{\the\subsectnumber}}.%
           {\the\subsubsectstyle{\the\subsubsectnumber}}\fi}%
           \wlog{\string\section\space \subsubsectlabel}%
       }%
       \the\subsubsectionformat{\the\subsubsectionfontstyle%
          \noindent{\ifcn@@\unskip\hskip\subsectionindent%
            \subsubsectlabel~~\fi#1}}%
       \par%
       \nobreak\vskip\headskip \penalty 30000%
       \ifmakename%
           \csnamech@ck%
           \ifcn@@%
              \immediate\write\csnamewrite{\def\the\tempnametoks{%
                {\the\chapterstyle{\the\chapternumber}.%
                 \the\sectionstyle{\the\sectionnumber}.%
                 \the\subsectionstyle{\the\subsectnumber}.%
                 \the\subsubsectstyle{\the\subsubsectnumber}}}%
              }%
            \fi%
            \global\makenamefalse%
       \fi%
       \ifcontentson%
          \c@ntentscheck%
          \CONTENTS{x};{#1}%
       \fi%
   }%

   \def\contentsinput{%
       \ifcontentson%
           \contentsopenfalse%
           \immediate\closeout\contentswrite%
           \global\oldheadline=\headline%
           \global\headline={\hfill}%
           \global\oldfootline=\footline%
           \global\footline={\hfill}%
           \fontsoff \unlock%
           \input \the\jobdir\jobname.contents%
           \fontson%
           \lock%
           \endpage%
           \global\headline=\oldheadline%
           \global\footline=\oldfootline%
       \else%
           \relax%
       \fi%
   }


       \def\phyzzxfootline{
           \footline={\ifletterstyle\the\letterfootline%
               \else\the\paperfootline\fi}%
       }

%

   {\obeyspaces}

   \def\verbfile#1{
       {\catcode`\\=12\catcode`\{=12
       \catcode`\}=12\catcode`\$=12\catcode`\&=12
       \catcode`\#=12\catcode`\%=12\catcode`\~=12
       \catcode`\_=12\catcode`\^=12\obeyspaces\obeylines\tt
       \verbdonetrue\openin\verbinfile=#1
       \loop\read\verbinfile to \inline
           \ifeof\verbinfile
               \verbdonefalse
           \else
              \leftline{\inline}
           \fi
       \ifverbdone\repeat
       \closein\verbinfile}
   }

   \def\boxit#1{\vbox{\hrule\hbox{\vrule\kern3pt%
       \vbox{\kern3pt#1\kern3pt}\kern3pt\vrule}\hrule}%
   }

   \def\square{%
      \setbox\squarebox=\boxit{\hbox{\phantom{x}}}
      \squareht = 1\ht\squarebox
      \squarewd = 1\wd\squarebox
      \vbox to 0pt{
          \offinterlineskip \kern -.9\squareht
          \hbox{\copy\squarebox \vrule width .2\squarewd height .8\squareht
              depth 0pt \hfill
          }
          \hbox{\kern .2\squarewd\vbox{%
            \hrule height .2\squarewd width \squarewd}
          }
          \vss
      }
   }

   \def\fboxit#1#2{
       \vbox{\hrule height #1
           \hbox{\vrule width #1
               \kern3pt \vbox{\kern3pt#2\kern3pt}\kern3pt \vrule width #1
           }
           \hrule height #1
       }
   }

   \let\eqnameold=\eqname

   \def\draft{\def\eqname##1{\eqnameold##1:{\tt\string##1}}
      \let\eqnalign = \eqname
   }
%
%
   \def\runningrightheadline{%
       \hfill%
       \tenit%
       \ifstartofchapter%
          \global\startofchapterfalse%
       \else%
          \ifcn@@ \the\chapternumber.\the\sectionnumber\quad\fi%
              {\fontsoff\thesectionhead}%
       \fi%
       \qquad\twelverm\folio%
   }

   \def\runningleftheadline{%
      \twelverm\folio\qquad%
      \tenit%
      \ifstartofchapter%
          \global\startofchapterfalse%
      \else%
         \ifcn@@%
             Chapter \the\chapternumber \quad%
         \fi%
         {\fontsoff\thechapterhead}%
         \hfill%
      \fi%
   }

   \runningheadlines={%
      \ifodd\pageno%
         \runningrightheadline%
      \else%
         \runningleftheadline%
      \fi
   }

%
%
%
%
%

   \font\dfont=cmr10 scaled \magstep5


   \newbox\cstrutbox
   \newbox\dlbox
   \newbox\vsk

   \setbox\cstrutbox=\hbox{\vrule height10.5pt depth3.5pt width\z@}

   \def\cstrut{\relax\ifmmode\copy\cstrutbox\else\unhcopy\cstrutbox\fi}

   \def\dl #1{\noindent\strut
       \setbox\dlbox=\hbox{\dfont #1\kern 2pt}%
       \setbox\vsk=\hbox{(}%
       \hangindent=1.1\wd\dlbox
       \hangafter=-2
       \strut\hbox to 0pt{\hss\vbox to 0pt{%
         \vskip-.75\ht\vsk\box\dlbox\vss}}%
       \noindent
   }

%
%

   \newdimen\fullhsize

   \fullhsize=6.5in
   \def\fullline{\hbox to\fullhsize}
   \let\l@r=L

   \newbox\leftcolumn
   \newbox\midcolumn

   \def\twocols{\hsize = 3.1in%
%
%
%
%
%
      \doublecolskip=.3333em plus .3333em minus .1em
      \global\spaceskip=\doublecolskip%
      \global\hyphenpenalty=0
      \singlespace
      \gdef\makeheadline{%
          \vbox to 0pt{ \skip@=\topskip%
          \advance\skip@ by -12pt \advance\skip@ by -2\normalbaselineskip%
          \vskip\skip@%
          \fullline{\vbox to 12pt{}\the\headline}\vss}\nointerlineskip%
      }%
      \def\makefootline{\baselineskip = 1.5\normalbaselineskip
           \fullline{\the\footline}
      }
      \output={%
          \if L\l@r%
             \global\setbox\leftcolumn=\columnbox \global\let\l@r=R%
          \else%
              \doubleformat \global\let\l@r=L%
          \fi%
          \ifnum\outputpenalty>-20000 \else\dosupereject\fi%
      }
      \def\doubleformat{
          \shipout\vbox{%
             \makeheadline%
             \fullline{\box\leftcolumn\hfil\columnbox}%
             \makefootline%
          }%
          \advancepageno%
      }
      \def\columnbox{\leftline{\pagebody}}
      \outer\def\twobye{%
          \par\vfill\supereject\if R\l@r \null\vfill\eject\fi\end%
      }%
   }

   \def\threecols{
       \hsize = 2.0in \tenpoint

      \doublecolskip=.3333em plus .3333em minus .1em
      \global\spaceskip=\doublecolskip%
      \global\hyphenpenalty=0

       \singlespace

       \def\makeheadline{\vbox to 0pt{ \skip@=\topskip
           \advance\skip@ by -12pt \advance\skip@ by -2\normalbaselineskip
           \vskip\skip@ \fullline{\vbox to 12pt{}\the\headline} \vss
           }\nointerlineskip
       }
       \def\makefootline{\baselineskip = 1.5\normalbaselineskip
                 \fullline{\the\footline}
       }

       \output={
          \if L\l@r
             \global\setbox\leftcolumn=\columnbox \global\let\l@r=M
          \else \if M\l@r
                   \global\setbox\midcolumn=\columnbox
                   \global\let\l@r=R
                \else \tripleformat \global\let\l@r=L
                \fi
          \fi
          \ifnum\outputpenalty>-20000 \else\dosupereject\fi
       }

       \def\tripleformat{
           \shipout\vbox{
               \makeheadline
               \fullline{\box\leftcolumn\hfil\box\midcolumn\hfil\columnbox}
               \makefootline
           }
           \advancepageno
       }

       \def\columnbox{\leftline{\pagebody}}

       \outer\def\threebye{
           \par\vfill\supereject
           \if R\l@r \null\vfill\eject\fi
           \end
       }
   }


%
%
%

   \everyjob{%
      \xdef\today{\monthname~\number\day, \number\year}
      
       \immediate\openin\testifexists=m
       \ifeof\testifexists
           \immediate\closein\testifexists
       \else
         \immediate\closein\testifexists
         \input m
       \fi
   yphyx.tex
      \ifforwardrefson%
         
       \immediate\openin\testifexists=\the
       \ifeof\testifexists
           \immediate\closein\testifexists
       \else
         \immediate\closein\testifexists
         \input \the
       \fi
   \jobdir\jobname.csnames
      \fi%
   }

\contentsoff
\lock
\catcode`\@=12 

\topskip=0pt
\normalparskip=0pt plus 0.1pt minus 0.1pt
\def\normalrefmark#1{$^{\scriptstyle #1}$}
\refindent=19pt
\unlock
\def\refitem#1{\r@fitem{#1.}}
\def\foottextindent#1{\indent\llap{#1\enspace}\ignorespaces}
\def\Vfootnote#1{\insert\footins\bgroup
   \interlinepenalty=\interfootnotelinepenalty \floatingpenalty=20000
   \singl@true\doubl@false\Tenpoint
   \splittopskip=\ht\strutbox \boxmaxdepth=\dp\strutbox
   \spaceskip=\z@skip \xspaceskip=\z@skip \footnotespecial
   \foottextindent{#1}\footstrut\futurelet\next\fo@t}
\lock
\def\SCIPP{\centerline{Santa Cruz Institute for Particle Physics}
  \centerline{University of California, Santa Cruz, CA 95064}}

\def\unlock{\catcode`@=11} 
\def\lock{\catcode`@=12} 
\unlock
\def\normalrefmark#1{$^{\scriptstyle #1}$}
\def\refitem#1{\r@fitem{#1.}}
\lock
\def\SCIPP{\centerline{Santa Cruz Institute for Particle Physics}
  \centerline{University of California, Santa Cruz, CA 95064}}
\Pubnum{SCIPP 92/36}
\date{September 1992}
\pubtype{ T}
\titlepage
\vskip3cm
\centerline{\fourteenbf STRING THEORY:}
\centerline{\fourteenbf LESSONS FOR LOW ENERGY PHYSICS?}
\vskip12pt
\centerline{{Michael Dine}
\foot{Work supported in part by the U.S.~Department of Energy.}}
\vskip2pt
\SCIPP
\vfill

\singlespace
\centerline{Abstract}

\vskip5pt\noindent
This talk considers possible lessons of string theory for low energy
physics.  These are of two types.  First, assuming that string
theory is the correct underlying theory of all interactions,
we ask whether there are any generic predictions the theory makes,
and we compare the predictions of string theory with those of
conventional grand unified theories.  Second, string theory
offers some possible answers to a number of troubling naturalness
questions.  These include problems of discrete and continuous
symmetries in general, and CP and the strong CP problem in
particular.

\vfill
\centerline {Invited Talk}
\centerline{International Conference on High Energy Physics}
\centerline{Dallas, Texas, August, 1992.}
\vfill
\endpage

I have been asked by the organizers to discuss lessons that
string theory might hold for low energy physics.  This is a
difficult charge.  Some would argue that it
is likely that string theory is the
underlying theory of all interactions.  In this view, we should
simply wait until we understand how to connect
the theory to reality.  Others argue that
the theory might well have
nothing to do with nature, or -- perhaps worse -- that while it may be
the ``theory of everything" it might take millennia
to connect it to reality.
Most who hold these views believe that the theory is unlikely to
teach us anything.  In this lecture, I would like
to adopt a middle ground.  My own  opinion is that
string theory, whether
or not it is ultimately correct, has a number of interesting lessons
to offer us.  After all, it is, at least potentially, a truly unified
theory, and thus it is possible to ask of it
many of the questions which trouble us in field theory.  As we
will see, string theory has already provided interesting
answers to several
problems of symmetries and naturalness; these will be
the subject of this talk.  Numerous other
important topics will not be covered here.
For example, at the present time, much of
the effort in string theory is
being devoted to questions such as the possibility of information
loss in black holes.  While these studies are very exciting, they
have not yet yielded definite conclusions.

In thinking about problems which trouble us in conventional
field theory model building, I believe it is reasonable to
assume that anything which happens in string theory can
occur in whatever may be the ultimate theory.  In this spirit,
I will attempt to
apply the general lessons string theory teaches about naturalness
to more conventional problems of low energy supersymmetry
and grand unification.  However,
apart from general observations about how this prototype unified
theory resolves (or fails to resolve) certain questions of naturalness,
we should also keep in mind the possibility that
string theory {\it is} the ultimate
theory of nature.  So it is also interesting to
ask whether string theory itself
makes any generic predictions, independent
of the details of compactification and the like.  Unfortunately, to
date the answer is no, but there are some features which are {\it almost}
generic.  I will mention some of these as I go along.
Of course, my choices here, as those above, reflect my
prejudices and interests.
There are many things which will not be covered
in any detail in this talk.
In particular, I will only briefly mention
\REF\kaplunovsky{N.V. Krasnikov, Phys. Lett. {\bf B293} (1987) 37.
L. Dixon, V. Kaplunovsky, J. Louis and M. Peskin, unpublished;
L. Dixon, {\it Proceedings of the 1990 DPF Meeting}, World Scientific,
Singapore, 1990; J. Louis, {\it Proceedings of the 1991 DPF Meeting},
World Scientific, Singapore, 1991.}
some of the work which has gone into developing a theory of
supersymmetry-breaking in string theory.\refmark{\kaplunovsky}
I will not have time to
discuss efforts to develop a detailed phenomenology based on
particular string models.   Some of these will be presented by
other speakers at this meeting.
The reader should thus be forewarned.

\REF\gsw{M. Green, J. Schwarz and E. Witten, {\it Superstring Theory},
Cambridge University Press, New York, 1986.}
All of the discussion will be in one particular framework:  we will
study classical solutions of the string equations [(super-conformal
field theories] with four flat, Min\-kowski directions.  Within this
framework, string theory has scored a number of impressive successes,
and it is worth listing them.\refmark{\gsw}  One finds:
\item{1.}  Chiral fermions (generations)
\item{2.}  Low energy supersymmetry
\item{3.}  Axions
\item{4.}  Massless Higgs doublets unaccompanied by triplets
(more generally, particles which are massless which are permitted
to gain mass by all space-time symmetries)
\item{5.}  A rich structure of discrete symmetries
\item{6.}  Gravity!

Items 3 and 4 already represent significant violations of
conventional field-theoretic
\REF\miraculous{M. Dine and N. Seiberg, Nucl. Phys. {\bf B306} (1988)
137.}
notions of naturalness.  Some time ago, Nathan Seiberg and I suggested
that the term ``string miracle" should have a technical meaning, referring
precisely to phenomena of this kind.\refmark{\miraculous}
We will explain shortly in what
sense the string axion is miraculous, according to this definition.
Masslessness of Higgs doublets, in conventional grand unified
theories -- even
with supersymmetry -- requires fine tuning.\foot{To the best of my
knowledge, various proposals which have been offered to explain
massless Higgs doublets in field theory fail once one considers
higher dimension operators.}

String vacua have other remarkable features.  Not only do they
exhibit generations, for example, but in any given vacuum one can
calculate (sometimes easily, sometimes with substantial labor)
Yukawa couplings.  So string theory does truly have pretensions to
be a complete unified theory.

\REF\rohm{R. Rohm, Nucl. Phys. {\bf B237} (1984) 553.}
Unfortunately, against these successes one must weigh some serious --
perhaps catastrophic -- failures.  I will list three here; some other
potential problems will be considered later.  By far the most serious
problem is the cosmological constant problem; string theory has offered
no insight into the question of why the cosmological constant vanishes.
Specifically, whenever supersymmetry is broken and one can calculate
the cosmological constant, it is non-zero and its magnitude agrees
with naive estimates (\ie, it is some power of the SUSY breaking scale
times a suitable power of a cutoff).\refmark{\rohm}

\REF\strongcoupling{M. Dine and N. Seiberg, Phys. Lett. {\bf 162B}
(1985) 299.}
\REF\dixonetal{See, \eg, L. Dixon, V. Kaplunovsky and J. Louis,
Nucl. Phys. {\bf B355} (1991) 649; Nucl. Phys. {\bf B329} (1990) 27.}
Closely related to this question is the
problem that, while string vacua have many attractive features, there
are far, far too many of them.  There are two senses in which this number
is large:  there are many discrete choices of solutions (characterized,
\eg, by the dimensionality of space-time, the number of generations,
and similar quantities), and there are continuous parameters.  Among
the latter are the value of the coupling constant, and the size and
shape of the internal spaces.  All of these are determined by the
expectation values of dynamical fields; at least in perturbation theory
these fields have no potential (in the case that supersymmetry is unbroken)
and determining their values is part of the problem of supersymmetry
breaking.  Even without knowledge of the mechanism of supersymmetry
breaking, however, one can already see that we are headed for
trouble:\refmark{\strongcoupling}
necessarily, the potential for the field which determines the coupling --
the ``dilaton" -- vanishes at zero coupling, so the potential will
always have a minimum where the theory is free.  Any other minimum
is likely to lie at strong coupling, where perturbation theory
is not useful.   There have been some interesting proposals for
solving this problem.\refmark{\kaplunovsky}  In addition, a great
deal of machinery has been developed for dealing with the problem
of supersymmetry breaking.\refmark{\dixonetal}  But it is probably fair
to say that to date no compelling picture for supersymmetry breaking
has been offered.

As we proceed, we will see that there are
other problems.  Perhaps these are somehow overcome; perhaps not.
But even if string theory is not the ultimate theory of nature, it
does provide an interesting framework -- at the moment possibly the only
framework -- to address questions at a variety of scales which trouble
particle physicists.
The remainder of this talk,
as suggested by our remarks above, has three themes:
\item{1.}  We will
ask whether there are any generic string predictions.  While we
cannot give a definitive answer, we will point out a number of
features common to many string compactifications, which might have
implications for low energy phenomenology.
\item{2.}  We will consider the distinctions between string theory and
more general field theories.
\item{3.}  We will consider a number of questions of naturalness
which have been raised in field theory model building, and see
how they are resolved in string theory.  Viewing string theory as
a paradigm for unification, we will suggest that phenomena which occur in
string theory might plausibly occur in any ultimate theory, and
thus these observations can (and will) serve as a guide to model building.

In the next section, we will compare string theory with conventional
grand unified theories.  We will also note that string theory seems unlikely
to yield something like the minimal supersymmetric standard model:
there are likely to be gauge singlets and R-parity
is likely broken.  We will point out that
string theory provides
no magic answers to the problems of flavor-changing processes in these
theories.
In the third section we will discuss symmetries.  We will see that string
theory often produces approximate, global discrete symmetries.  In
the fourth section,  we will consider CP and the strong CP problem.
We note that CP is a gauge symmetry in string theory, which must be
spontaneously broken.   This breaking may occur near $M_p$, or
at lower
energies.  In the former case, axions are probably required to solve the
strong CP problem, and we consider some aspects of axions in string theory.  In
the case of lower energy breaking, it is natural to
consider
other possible solutions.  Within the framework of low energy supersymmetry,
this leads to general
predictions about the form
of CP violation at low energies.

\vskip15pt
\centerline{STRING THEORY, GRAND UNIFICATION}
\centerline{AND LOW ENERGY SUPERSYMMETRY}
\vskip7pt
\REF\couplings{V. Kaplunovsky, Phys. Rev. Lett. {\bf 55} (1036) 1985.}
\REF\bfms{T. Banks, L. Dixon, D. Friedan and E. Martinec, Nucl.
Phys. {\bf B299} (1988) 613.}
To begin, it is worthwhile to make some comparisons between string
theory and conventional grand unified theories.  One of the
striking features of string theory is that the gauge couplings
are all equal at the fundamental scale of the theory\refmark{\couplings}
(which
we can loosely think of as the string scale of the Planck mass).
In other words, they are precisely equal at the tree level.  From
a field theory point of view, this is quite amazing.  In a conventional
grand unified theory, higher dimension operators can
correct such relations (albeit by a small amount).  For example,
in SU(5), if $\Sigma$ is a 24 with a
non-zero vev, higher dimension operators such as
$1/M_p^2 {\rm Tr} \Sigma^2
F^2$ will break the equality between the couplings.  This
illustrates that the effective field theory which describes strings,
while it contains operators of arbitrarily high dimension (already
at tree level), is a very special one.  It is also true that
under quite general circumstances, $\sin^2(\theta_W)=3/8$,
exactly, at tree level.\refmark{\bfms}

\REF\stringunity{See, for example,
G.G. Ross and R.G. Roberts, preprint RAL-92-005; M. Gaillard and R. Xiu,
preprint LBL-32324 (1992).}
While remarkable, these statements have an unfortunate consequence
for string phenomenology.  There has been much excitement recently
over the fact that the minimal supersymmetric standard model
(MSSM) leads naturally to unification of coupling constants with
a unification scale of order $10^{16}$ GeV.  In string theory,
on the other hand, one can give good arguments that unification occurs
very near $M_p$.\refmark{\couplings}
Thus string unification requires that
one have something beyond the minimal standard model:  additional
particles and interactions, some early partial unification, or
something else.\refmark{\stringunity}
This is not a disaster, but it is disappointing
that things do not work simply.

\REF\banksdine{T. Banks and M. Dine, Phys. Rev. {\bf D45} (1992)
424.}
Returning to the comparison of strings and GUT's, there are some
striking differences.  First,  GUT type relations
among Yukawa couplings do not hold.\refmark{\gsw}  In addition,
if one examines the transformation laws of the light fields under
discrete symmetries, one finds,
in general, that they are not related as they would be in
a grand unified theory.  Typically in a field theory
the light fields would consist of complete multiplets
of the unified gauge group [\eg, the $\bar 5$ and $10$
of SU(5)], each transforming in the same representation
of the discrete symmetry.\foot{One can construct
counterexamples to this statement in the
following way.  Suppose that the discrete symmetry
which survives to low energies is a linear combination
of the original discrete symmetry and a gauge transformation
of the unified theory.  Then different elements of a multiplet
will transform differently.  This might
give a framework in which to explain the
existence of massless doublets, though I have not succeeded in
doing this.}
There may
also be incomplete mirror multiplets
(\eg, the Higgs doublets in a supersymmetric SU(5) theory);
these will also be mirrors with respect to the discrete symmetry,
\ie, they will have quantum numbers which permit a mass term.
In typical string models, on the other hand, the discrete quantum
number assignments look almost random (consider the models
discussed in ref.~\gsw, adding Wilson lines).
They seem subject only to very mild constraints.\refmark{\banksdine}

\REF\singletproblems{J. Polchinski and L. Susskind, Phys. Rev. {\bf D36}
(1982) 3661; M. Dine and W. Fischler, Nucl. Phys. {\bf B204}
(1982) 346.}
Both of these observations suggest
strategies for model building, which have already been
exploited in various string-inspired models.
It is worth mentioning some other features of string compactifications
with implications for model building.  One is that these
models typically predict the existence of light gauge singlets.
It is probably not reasonable to call this a string prediction;
I know of no theorem that these fields need arise.  However,
they are extremely common, and this suggests one should seriously
consider extensions of the minimal theory with such fields.
\foot{Note here that I am not referring to the moduli of $(2,2)$
compactifications, but rather, for example, to the O(10) singlets
which arise in $E_6$ and which can couple to Higgs doublets.}
Such theories  suffer from well-known
problems\refmark{\singletproblems}, such as appearance of large
tadpoles,
but these can readily be
solved with discrete symmetries.

\REF\rparity{L. Hall and M. Suzuki, Nucl. Phys. {\bf B231} (1984) 419;
S. Dimopoulos and L. Hall, Phys. Lett. {\bf 207B} (1988) 210.}
A second point concerns $R$-parity violation.  The discrete
symmetries which tend to arise at low energies are often
rather intricate.  If string theory, or something like it, is
the underlying theory, there is no reason to think that
the symmetry which forbids proton decay is the simplest
$R$-parity.  This suggests one should take very seriously
the possibility of $R$-parity violation.\refmark{\rparity}

\REF\vadimfcnc{V. Kaplunovsky, Texas preprint in preparation.}
Since we are on the subject of low energy supersymmetry, there is
another lesson which string theory teaches.  It is well known
that absence of flavor changing neutral currents in supersymmetry
requires an approximate degeneracy of squark masses.  Such a
degeneracy is not enforced by any symmetry.
In hidden sector supergravity
models, for example, if $Z$ is a hidden sector field, and
$\phi_i$ are observable fields, terms in the Kahler potential
of the form
$$\gamma_{ij} Z^{\dagger} Z \phi^{i~\dagger} \phi^j\eqn\fbreaking$$
contribute to scalar masses.  Mass degeneracy requires that
$\gamma$ should be proportional to the unit matrix.  In a generic
string vacuum, there is no reason to expect such a feature.
Kaplunovsky\refmark{\vadimfcnc} has calculated many of these
couplings and shown, indeed, that nothing of this sort
happens.

\REF\nanop{J. Ellis, C. Kounnas, and D.V. Nanopoulos, Nucl. Phys.
{\bf B247} (1984) 373; A.B. Lahanas and D.V. Nanopoulos,
Phys. Rep. {bf 145} (1987) 1.}
\REF\dks{M. Dine, A. Kagan and S. Samuel, Phys. Lett. {\bf 243B}
 (1990) 250.}
\REF\oldmd{M. Dine, Nucl. Phys. {\bf B224} (1983) 93.}
That there is no generic solution does not mean that there do not
exist solutions.  In the context of string inspired models,
there has been one interesting suggestion:  perhaps at the large
scale, the scalar masses are essentially zero, while gaugino masses
are not.\refmark{\nanop}  In ref.~\dks, it is shown that with
some reasonable naturalness constraints, one can obtain in this
way a low energy spectrum with adequate degeneracy.\foot{Apart
from the imaginary part of the $K$-$\bar K$ mass matrix.  This
requires that some phases be small, as do the neutron and electron
electric dipole moments.
We will say more about CP later.}
Alternatively, the problem can be solved if supersymmetry breaking
is communicated from the hidden sector primarily by gauge interactions
rather than by gravity.\refmark{\oldmd}

\vskip15pt
\centerline{SYMMETRIES}
\vskip7pt
\REF\banksdixon{T. Banks and L. Dixon, Nucl. Phys. {\bf B307} (1988)
93.}
In the context of model building in field theory, it has long
been argued that global symmetries are unnatural.  Apart
from aesthetic considerations, it is not clear how such symmetries
could survive gravitational corrections.  String theory lends
support to this view:  one can prove, without much difficulty and
with very weak assumptions, that there are no
(unbroken) global continuous
symmetries in string theory; all continuous symmetries are
gauge symmetries.\refmark{\banksdixon}

\REF\dinecp{M. Dine, R. Leigh and D. MacIntire, to appear in
Phys. Rev. Lett., SCIPP Preprint 92/16  (1992).}
We have already remarked that discrete symmetries often arise
in string theory.  Frequently, one can think of these as gauge symmetries,
or as general coordinate transformations in some higher dimensional
space.  In other words, they are surviving subgroups of some
larger (continuous) gauge symmetry group.  Indeed, it is
widely believed
that {\it all} discrete symmetries in string theory are of this
kind, but this statement is difficult to prove.  It is amusing that
one {\it can} show that the $Z_2$ symmetry which exchanges the
two $E_8$'s of the heterotic string theory is a gauge
symmetry.\refmark{\dinecp}

\REF\banksdiscrete{T. Banks, Physicalia {\bf 13} (1990) 19.}
\REF\krausswilczek{L. Krauss and F. Wilczek, Phys. Rev. Lett. {\bf
62}, 1221 (1989).}
\REF\ir{L. Ibanez and G. Ross, Phys. Lett. {\bf 260B} (1991) 291;
CERN-TH-6000-91 (1991).}
Independently of string theory, it has been argued that global discrete
symmetries are likely to be broken by gravity.\refmark{\banksdiscrete}
On the other hand, gauged discrete symmetries of the type discussed
above almost surely survive.\refmark\krausswilczek\
These considerations led Ibanez and
Ross\refmark\ir\
to ask what consistency conditions might be required of a low
energy theory in order that a discrete symmetry could be gauged.
To address this question, they embedded the symmetry in a continuous
group, and examined the anomaly cancellation conditions for this
larger group.  To derive conditions in this way requires making some
assumptions about charges of heavy fields.
In ref.~\banksdine, T. Banks and I pointed out that many string
compactifications violate the simplest set of assumptions.  The only
constraints which hold independent of such assumptions
can be understood in terms of instantons of the low energy theory.

In the course of this work, however, we made another discovery.
In numerous ground states of string theory, one finds discrete
symmetries with anomalies.  One might try to conclude from this
that either the corresponding symmetries are not truly gauge
symmetries, or that the corresponding compactifications are
inconsistent.  However, in all of the cases which have been
examined to date, it is possible to cancel the anomaly by
assigning to the so-called model-independent axion a non-linear
transformation law under the discrete symmetry.  In other words,
an instanton generates a non-zero amplitude of the type
$$\psi \psi \psi \dots \psi e^{-8 \pi^2/g^2} e^{i a/f_A}
\eqn\instantonamplitude$$
where $a$ is the axion, and $\psi$ denotes some space-time fermion
field.  If one ignores the instanton, the combination of fermions
here (by assumption) violates the discrete symmetry, say by
$e^{i \Delta}$.  However,
by assuming $a \rightarrow a- \Delta$ under the symmetry, the
amplitude becomes invariant.  Of course, this means that the
non-anomalous discrete symmetry is spontaneously broken.  What
is much more interesting, however, is
that in perturbation theory there is nothing
wrong with the original (unbroken) symmetry.  Only non-perturbatively
is there any violation of the symmetry.  This suggests that it
is reasonable to consider global discrete symmetries, which are only
broken by small amounts.  Such a possibility is of interest for
questions such as fermion mass hierarchies and flavor changing
neutral currents, as well as for suppressing baryon violation
and other unwanted processes in supersymmetric models.

\vskip15pt
\centerline{CP, STRONG CP, AND ALL THAT}
\vskip7pt

\REF\ggw{H. Georgi, S. Glashow and M. Wise, Phys. Rev. Lett.
{\bf 47} (1981) 402.}
\REF\march{M. Kamionkowski and J. March-Russell, Phys. Lett.
{\bf 282B} (1992) 137;  R.~Holman {\it et al.},
Phys. Lett. {\bf 282B} (1992) 132;
S.M. Barr and D. Seckel, Bartol preprint BA-92-11.}
In field theory, the axion solution of the strong CP problem
begins by postulating that classically and in perturbation theory,
nature possesses a Peccei-Quinn symmetry,
$$a(x) \rightarrow a(x) + \delta\ . \eqn\pq$$
This symmetry is then broken non-perturbatively by
QCD effects.  From a field-theoretic perspective
this may be ``natural" in the sense that, given the symmetry,
the corrections are small; still,
it is disturbing to postulate such
a broken symmetry.\refmark{\ggw}
As has been stressed recently, however,
if gravity violates global symmetries, it is almost impossible
to understand how the axion solution can possibly work.\refmark{\march}
Suppose
that the Peccei-Quinn symmetry is broken by a high-dimension coupling,
$${\cal L}_{SB}={\gamma \over M_p^n}{\cal O}_{n+4}\eqn\symmbreak$$
where ${\cal O}_{n+4}$ denotes an operator of dimension $n+4$,
and $\gamma$ is a numerical constant, presumably of order one.
Such a coupling will generate a linear term in the axion potential,
with coefficient of order
$${\gamma f_a^{n+3} \over M_p^n} a(x)\ .\eqn\linearterm$$
If $f_a \sim 10^{11}$ GeV, then requiring $\theta < 10^{-9}$ means that
one must have $n \ge 8$, \ie, one must suppress all potential
symmetry breaking operators up to dimension $12$.  If $f_a =M_p$,
one needs to suppress all possible symmetry-breaking operators.

{}From this perspective, it should be clear why I referred to the
presence of axions in string theory as a ``miracle."  In compactifications
of the heterotic string, there is always at least one, ``model-independent"
axion.  The existence of this axion can be understood in at least
two ways.  First, the low energy theory always contains a two-index
antisymmetric tensor field.  This field is equivalent to a scalar.
The couplings of the antisymmetric tensor are governed by a gauge
principle; this principle can be shown to forbid a mass term
for the scalar.  Alternatively, if one studies the vertex operator
for this axion, it is readily seen to be (from a world-sheet viewpoint)
a total divergence at zero momentum, indicating that the axion
has only derivative couplings.  It is also not hard to show that
this field has the correct $F \tilde F$ couplings.

\REF\nsh{J. Preskill, M. Wise and F. Wilczek,
Phys. Lett. {\bf 120B}(1983) 127; L. Abbott and P. Sikivie,
Phys. Lett. {\bf 120B} (1983) 133; M. Dine and W. Fischler,
Phys. Lett. {\bf 120B} (1983) 137.}
\REF\loopholes{J. Cline and S. Raby, Phys. Rev. {\bf D43} (1991) 1381;
A.D. Linde, Phys. Lett. {\bf 259B} (1991) 38.}
While the presence of this axion is ``miraculous," there are two
reasons why it might not solve the strong CP problem.  First,
in many string models, there are strong gauge groups besides
SU(3).  Thus one may need additional axions.  Second, the large
decay
constant means that it violates the cosmological limit,
$f_a < 10^{12}$ GeV.\refmark{\nsh} It is not clear how much one
should trust this limit; various loopholes have been suggested through
the years.\refmark{\loopholes}

\REF\lazarides{G. Lazarides, C. Panagiotakopoulos and Q. Shafi,
Phys. Rev. Lett. {\bf 56} (1986) 432.}
\REF\casas{J. Casas and G. Ross, Phys. Lett. {\bf 192B} (1987) 119.}
\REF\mangano{M. Dine, V. Kaplunovsky, M. Mangano, C. Nappi and N.
Seiberg, Nucl. Phys. {\bf B259} (1985) 549.}
\REF\lou{M. Dine, to be published in {\it Proceedings of the
Cincinnati Symposium in Honor
of the Retirement of Louis Witten}, SCIPP Preprint 92/27 (1992).}
However, for now, let us take
the limit seriously, and ask how one might obtain such an axion.
In the past, a number of authors have considered the possibility
that the Peccei-Quinn symmetry might arise accidentally, as a consequence
of properties of the low dimension operators in some effective field
theory.  In the context of string theory, at least two sets of authors
have suggested that discrete symmetries might lead to such an
approximate symmetry.\refmark{\lazarides,\casas}
The authors of ref.~\casas\ attempted
to estimate the effective $\theta$ which would appear in their
model, precisely along the lines discussed above.
One extremely nice feature of the string-inspired models considered
by these authors is that the axion decay constant is automatically
of order\refmark{\mangano}
$M_{INT} \sim \sqrt{M_W M_p}
\sim10^{10}$--$10^{12}$ GeV, \ie, within the allowed
axion window.
The scale of the vev's which break the PQ symmetry is indeed
typically of this order.  Various aspects of these ideas, including
certain pitfalls not noted in earlier work, are discussed in ref.~\lou.

\REF\dineseibergcp{M. Dine and N. Seiberg, Nucl. Phys. {\bf B273}
(1986) 109.}
\REF\shenker{S. Shenker, Rutgers preprint RU-90-47 (1990).}
Even if string theory produces an axion in this way, there
is another concern:  the axion may fix $\theta$ to a
CP-violating value.
It is not necessarily true that an axion in string theory (or field
theory) solves the strong CP problem.
There is no guarantee, in general, that effects in the high
energy theory don't give rise to contributions to the axion potential
which are larger than the QCD contributions, particularly in theories
in which the QCD $\beta$-function is small or positive at high
energies.\refmark{\dineseibergcp}  In such cases, if
the high energy theory is not
CP-conserving, there is no reason for the
minimum of the axion potential to lie at the CP-conserving
point, so the Peccei-Quinn solution can be spoiled.
In ref.~\dineseibergcp, it was shown that small instantons
can lead to precisely this effect, and that this might occur
in many string compactifications.
Perhaps even
more worrisome is the observation of Shenker
that there may be effects in string theory which behave as $e^{-1/g}$,
\ie, which are far
larger than non-perturbative effects in the low
energy field theory.\refmark{\shenker}
If present, these could easily dwarf the QCD contribution
to the axion potential.\foot{This problem is connected to the observations of
ref.~\march, in the case of accidental axions with
decay constant
less than $M_p$.  In that case, small instantons generate effective
interactions which violate the would-be symmetry.  This can only occur
if the operators of interest violate the supposed discrete symmetry.
This, in turn, means that the discrete symmetry is anomalous in the sense
described above.   The coefficients of the symmetry violating operators
are exponentially small, in this case; the point here is that this may not
be small enough.}
All of this suggests
that one should consider theories where CP is
unbroken at $M_p$, and spontaneously broken by other
fields at much lower energy.

\REF\nelson{A. Nelson, Phys. Lett. {\bf 136B} (1984) 387.}
\REF\barr{S.M. Barr, Phys. Rev. Lett. {\bf 53} (1984) 329.}
In conventional model building,  an alternative to the axion
idea has been considered from time to time.  This is the possibility
that CP is an exact symmetry of the underlying lagrangian, and
that the ``bare" $\theta$ consequently vanishes.  This, by itself,
is not enough to insure that the observed $\theta_{\rm QCD}$
is small enough.
Nelson\refmark{\nelson} has proposed a scenario for obtaining sufficiently
small $\theta$; this scheme has been further developed by
Barr.\refmark{\barr}

\REF\ckncp{K. Choi, D. Kaplan and A. Nelson, UCSD preprint PTH 92-11
(1992).}
\REF\sw{A. Strominger and E. Witten, Comm. Math. Phys. {\bf
101} (341) 1985.}
Remarkably, one can show that in string theory, what we refer
to in four dimensions as CP is a gauge symmetry!\refmark{\ckncp,
\dinecp}  It is a combination of
a general coordinate transformation in the ten-dimensional space
and a gauge transformation in O(32) or $E_8 \times E_8$.
One consequence of this observation is that string theory cannot
possess bare $\theta$-parameters; $\theta_{\rm QCD}$ is ``calculable"
in this sense.  CP {\it can} be spontaneously broken in two ways:
\item{a.}  As Strominger and Witten noted some time ago,\refmark{\sw}
string compactifications typically contain CP odd ``moduli" (fields
with no potential; one can think of their expectation values
as determining
the size and shapes of the internal spaces in these theories).
Expectation values for these fields break CP.  Such vev's
correspond to breaking of CP near $M_p$.  It is hard to see
how $\theta_{\rm QCD}$ could turn out to be small under these circumstances,
unless there are axions.  One will still have to worry about
the problems described above.
\item{b.}  CP can be broken by expectation values for
CP-odd matter fields.  Apart from other virtues which we
will describe below, this has the feature that unknown high
energy effects won't spoil the Peccei-Quinn solution.

Recently, we have been considering a scenario of the second type.
We have constructed a number of ``string-inspired models"
in which a Nelson-Bar type mechanism is operative.
Consider unification in the group $E_6$ (of the type
suggested by the simplest
Calabi-Yau compactifications).\refmark{\gsw}
Matter fields fall in $27$'s and $\overline{27}$'s of $E_6$.  Under
the usual SO(10) gauge group, the $27$ decomposes as
$$27=16 + 10 + 1\ .\eqn\otendecomp$$
Of particular interest two us are the two standard model singlets
in this decomposition.  Apart from the O(10) singlet, which
we will denote by $S$, the $16$ contains a particle which can
be identified as the right-handed neutrino, which we will denote
${\cal N}$.  The $10$ contains two colored fields with the quantum
numbers of the $\bar d$ quark and its antiparticle; we denote
these by $\bar q$ and $q$, respectively.  If one assumes that
the low energy theory contains soft supersymmetry breaking terms
of the usual type, then it is natural, as above,
to obtain vev's for some
of the $S$ and ${\cal N}$ fields of order $10^{10}$ GeV,
where the $S$ vev's are real while the ${\cal N}$ vev's are
complex and CP violating.  This gives rise to a mass matrix
of the form discussed by Nelson and Barr:
$$\VEV{S} q \bar q + \VEV{\cal N} q \bar d\eqn\nbmassmatrix$$
As these authors have pointed out, this leads to a quark
mass matrix which is complex, but whose determinant is real.
Judicious choice of discrete symmetries insures that other
potential sources of $\theta$ are sufficiently suppressed.
For example, they insure reality of the Higgs mass matrix.
They can also assure other phenomenologically important properties,
such as suppression of $B$ and $L$ violation.

\REF\inprep{M. Dine, R. Leigh and D. MacIntire, SCIPP preprint
in preparation.}
Particular models of this type are described in ref.~\inprep.
They tend to be rather artificial-looking.  However, they all make
certain predictions:  they predict that the only source of CP
violation at low angles is the KM phase; there are no
additional phases in gaugino mass matrices, etc., and $\theta$
is extremely small.

\vskip15pt
\centerline{CONCLUSIONS}
\vskip7pt
My own view, from these observations, is that we have learned
some things from string theory, mostly about questions of naturalness.
Among these:
\item{1.}  While exact global continuous symmetries are unnatural,
gauged discrete symmetries are quite common; moreover,  it is
reasonable to postulate
weakly broken global discrete symmetries.
\item{2.}  It is plausible that if nature is supersymmetric at
low energies, it has a more complicated structure than that of the
MSSM.
\item{3.}  There is no obvious reason to think that, if nature is
supersymmetric, there should be any approximate flavor symmetry
of squark masses at the highest scales.  It is still necessary
to search for particular mechanisms which might explain the smallness
of flavor changing neutral currents.
\item{4.}  It is not unreasonable to postulate axions at very
high scales; axions at lower scales may arise accidentally as
consequences of discrete symmetries.
\item{5.}  Spontaneous CP-violation with small $\theta$ can arise as
a consequence of discrete symmetries.  This
may have other phenomenological consequences, such as vanishing
of CP-violating soft supersymmetry breaking phases.
\endpage
\refout
\end